\documentclass{article}
\usepackage[T1]{fontenc}
\usepackage[utf8]{inputenc}
\usepackage[english]{babel}
\usepackage{amsmath}
\usepackage{amssymb}
\usepackage{layouts}
\usepackage{graphicx}
\usepackage{enumerate}
\usepackage[normalem]{ ulem }
\usepackage{soul}
\usepackage[table]{xcolor}
    \definecolor{hotpink}{RGB}{255,105,180}
\usepackage[backend=biber,style=ieee]{biblatex}
\usepackage{booktabs}
\usepackage{csquotes}
\usepackage{subcaption}
\usepackage{systeme}
\usepackage{todonotes}
\usepackage[utf8]{inputenc}
\usepackage{tikz}
\usetikzlibrary{shapes.geometric, arrows}
\tikzstyle{startstop} = [rectangle, rounded corners, minimum width=3cm, minimum height=1cm,text centered, draw=black, fill=red!30]
\tikzstyle{io} = [trapezium, trapezium left angle=70, trapezium right angle=110, minimum width=3cm, minimum height=1cm, text centered, draw=black, fill=blue!30]
\tikzstyle{process} = [rectangle, minimum width=3cm, minimum height=1cm, text centered, draw=black, fill=orange!30]
\tikzstyle{decision} = [diamond, minimum width=3cm, minimum height=1cm, text centered, draw=black, fill=green!30]
\tikzstyle{arrow} = [thick,->,>=stealth]
\usepackage{comment}
\usepackage{geometry}
\usepackage{authblk}

\newcommand\Tstrut{\rule{0pt}{2.6ex}}         % = `top' strut
\newcommand\Bstrut{\rule[-0.9ex]{0pt}{0pt}}   % = `bottom' strut

\title{Method for estimating charge breeder ECR ion source plasma parameters with short pulse 1+ injection}

\author[1,$\dagger$]{J. Angot}
\author[2,*]{M. Luntinen}
\author[2]{T. Kalvas}
\author[2]{H. Koivisto}
\author[2]{R. Kronholm}
\author[4]{L. Maunoury}
\author[3]{O. Tarvainen}
\author[1]{T. Thuillier}
\author[2]{V. Toivanen}

\affil[1]{Univ. Grenoble Alpes, CNRS, Grenoble INP, LPSC-IN2P3, 53 Avenue des Martyrs, 38000 Grenoble, France}
\affil[2]{University of Jyväskylä, Department of Physics, Survontie 9D, 40500 Jyväskylä, Finland}
\affil[3]{STFC ISIS Pulsed Spallation Neutron and Muon Facility, Rutherford Appleton Laboratory, Harwell, OX11 0QX, UK}
\affil[4]{Grand Accélérateur National d’Ions Lourds, Boulevard Henri Becquerel, 14000 Caen, France}
\affil[$\dagger$]{julien.angot@lpsc.in2p3.fr}
\affil[*]{misapema@jyu.fi}

\date{December 2020}

\begin{document}

\maketitle

\begin{abstract}
    \centering
    A new method for determining plasma parameters from beam current transients resulting from short pulse 1+ injection into a Charge Breeder Electron Cyclotron Resonance Ion Source (CB-ECRIS) has been developed. The proposed method relies on few assumptions, and yields the ionisation times $1/n_e\left\langle\sigma v\right\rangle^{\text{inz}}_{q\to q+1}$, charge exchange times $1/n_0\left\langle\sigma v\right\rangle^{\text{cx}}_{q\to q-1}$, the ion confinement times $\tau^q$, as well as the plasma energy contents $n_e\left\langle E_e\right\rangle$ and the plasma triple products $n_e \left\langle E_e\right\rangle \tau^q$. The method is based on fitting the current balance equation on the extracted beam currents of high charge state ions, and using the fitting coefficients to determine the postdictions for the plasma parameters via an optimisation routine. 
    
    The method has been applied for the charge breeding of injected K$^+$ ions in helium plasma. It is shown that the confinement times of K$^{q+}$ charge states range from 2.6$^{+0.8}_{-0.4}$~ms to 16.4$^{+18.3}_{-6.8}$~ms increasing with the charge state. The ionisation and charge exchange times for the high charge state ions are 2.6$^{+0.5}_{-0.5}$~ms--12.6$^{+2.6}_{-3.2}$~ms and 3.7$^{+5.0}_{-1.6}$~ms--357.7$^{+406.7}_{-242.4}$~ms, respectively. The plasma energy content is found to be $2.5^{+4.3}_{-1.8}\times 10^{15}$~eV/cm$^3$.
 \end{abstract}
\section{Introduction}
Minimum-B Electron Cyclotron Resonance Ion Sources (ECRIS)~\cite{geller_ecris} are widely used to produce ion beams for particle accelerators and as charge breeders in ISOL facilities~\cite{geller06, blumenfeld13}. The performance improvements of these installations are dependent in part on the ECRIS development, which is driven by empirical laws obtained from decades of research and development. Despite recent studies using optical spectroscopy methods~\cite{kronholm19, kronholm20} and 1+/n+ diagnostics~\cite{tarvainen16, angot18}, understanding of the fundamental ECRIS plasma behavior, such as the ion confinement, the electron energy distribution or the plasma electrostatic potential distribution is still imprecise. 

In the ECRIS, Highly Charged Ions (HCIs) are created through stepwise electron-impact ionisation, whereas the principal charge decreasing mechanism is due to charge exchange with neutral atoms, which dominates over radiative recombination~(Ref.~\cite{mironov15} via~\cite{chung05}). The confinement of ions inside the plasma volume determines on one hand the time scale in which an ion may become further ionized, and on the other hand the rate at which the ions escape confinement and become available for beam formation. The plasma characteristic times --- the ion confinement time, the ionisation time and charge exchange time --- therefore play an important role in obtaining high intensity beams of HCIs. Shorter ionisation times and longer confinement and charge exchange times lead to more efficient HCI production. 

Experimentally, only a limited number of physical quantities can be observed, due in part to the fact that access to the plasma is technically challenging owing to the surrounding components, but also because the measurements must be limited to non-invasive methods in order to avoid corrupting the results by perturbing the plasma state. At present this means that one is limited to measuring the plasma radiation emissions, and the particles escaping confinement e.g. by measuring the extracted beam current. 

Transient methods for studying the effects of material injection on the extracted beam currents were first proposed by Pardo~\cite{pardo96}. These methods consist of pulsed injection of material into the support plasma and analysing the responses observed in the extracted current. Several techniques are used for pulsed material injection, e.g. laser ablation~\cite{harkewicz94}, fast gas injection~\cite{vondrasek02} sputtering~\cite{vondrasek02,neben16_experiments,marttinen20}, and 1+ injection~\cite{angot18}. A 0D code is then generally used to reproduce the measured beam currents by numerically optimising the plasma parameters involved in a system of differential equations formed by physical models describing the plasma~\cite{pardo96, mironov01, imanaka05}. 

In this work such transient measurements were conducted using short pulse 1+ injection of K into a He support plasma of a Charge Breeder ECRIS (CB-ECRIS)~\cite{angot_2020}. 1+ injection was chosen, because the number, energy and capture rate of injected particles can be precisely managed by tuning the 1+ source, contrary to e.g. sputtering where the yield depends on the energy of the ions bombarding the sample and, therefore, on the (unknown) local charge state distribution of the plasma. Alkali metal ions were chosen as the injected species to prevent wall recycling effects, and the helium buffer was used to obtain a clean Charge State Distribution (CSD) with as few overlapping peaks in the $q/m$ spectrum as possible. A critical reason for using K as the injected species was that the necessary ionisation rate coefficient data was readily available. 

We propose a new 0D-approach for analysing the current transients. For this purpose the balance equation governing the time evolution of the ion densities is converted into extraction current formalism, and the resultant differential equation is fitted on multiple consecutive charge state currents. The obtained fitting parameters are used to calculate the plasma characteristic times, energy content and triple product. The basis of the method is laid out in detail in sections~\ref{sec:theory}, and \ref{sec:numerical_method}, and the experimental methods are described in section~\ref{sec:experimental_methods}. As a further improvement to pre-existing 0D-methods, this one can be applied without a priori knowledge of input parameters ($T_e, T_i^q, n_e, n_0, \ldots$) or assumptions regarding the ion confinement scheme, while still including ionisation, charge-exchange and particle loss channels in the model. This method thus provides important improvements on the approach employed e.g. in Ref.~\cite{imanaka05}, where the ion beam current is taken to represent the plasma internal CSD directly, and the confinement time is assumed to be a linear function of the ion charge state $q$, which is not the case as per our results.

The scope of this work is to rigorously introduce the novel method which can then later be used for parametric studies, and to describe the experimental and numerical procedures involved. Access to the numerical code and sample data will be included as a supplementary material (see supplement~\ref{supplement:numerical_code} for a link to the repository).

\section{Theoretical foundations}\label{sec:theory}
The balance equation~\cite{shirkov91, melin99, geller_ecris} defines a group of coupled differential equations, which describe the time evolution of the ion population densities in the plasma. For the charge state $q$ the temporal change of the number density $n^q$\footnote{The subscript $i$ specifying the ion species is omitted as the support plasma perturbation is assumed small enough to be negligible, and the group of differential equations can be written solely for the injected species.} is given by:
\begin{equation}\label{eq:balance_equation}
    \begin{split}
    \frac{\mathrm{d}n^q}{\mathrm{d}t} = + \left\langle\sigma v\right\rangle^{\mathrm{inz}}_{q -1 \to q} n_e n^{q-1} &- \left\langle\sigma v\right\rangle^{\mathrm{inz}}_{q \to q+1} n_e n^{q} \\
    + \left\langle\sigma v\right\rangle^{\mathrm{cx}}_{q + 1\to q} n_0 n^{q+1} &- \left\langle\sigma v\right\rangle^{\mathrm{cx}}_{q\to q-1} n_0 n^{q}  \\
    &- \frac{n^q}{\tau^q}
    \end{split}
\end{equation}
Here the rate coefficients $\left\langle\sigma v\right\rangle^{\mathrm{inz}}_{q \to q+1}$ and $\left\langle\sigma v\right\rangle^{\mathrm{cx}}_{q \to q-1}$ describe ionisation from state $q$ to $q+1$, and charge exchange from state $q$ to $q-1$ respectively. The ionisation is assumed to occur in a stepwise process with electrons (density $n_e$) and charge exchange with the neutral atoms of the support plasma (density $n_0$). The last term in the balance equation depicts the actual loss rate of ions from the plasma volume, and defines the confinement time $\tau^q$ of the ion population. In our treatment both $n_e$ and $n_0$ are assumed to be fixed by the supporting He plasma.

The reaction rate coefficient is defined as the weighted average
\begin{equation}\label{eq:definition_rate_coeff}
    \left\langle\sigma v\right\rangle = 
    \int_{0}^{\infty} \sigma\left(v\right) v f\left(v\right) \mathrm{d}v,
\end{equation}
where $v$ is the relative speed of the interacting particles, $\sigma$ the reaction cross section, and $f$ the probability distribution for the relative speed. 

In this work, the ionisation rate coefficients of K$^{q+}$ were estimated using the semi-empirical expression by Voronov~\cite{voronov97}:
\begin{equation}\label{eq:voronov_rate_coeff}
    \left\langle\sigma v\right\rangle^{\mathrm{inz}}_{q \to q+1} = 
    A\frac{1 + P\cdot U^{1/2}}{X + U}U^K e^{-U} \quad \text{cm$^3$/s},
\end{equation}
with 
\begin{equation}
    U \equiv \frac{\delta E}{T_e}.
\end{equation}
Here the coefficients $A,P, K$ and $X$ are empirically determined coefficients from best fits to data, and $\delta E$ is the ionisation threshold energy. The coefficients are tabulated in Table~\ref{tab:voronov_coefficients} according to Ref.~\cite{voronov97}. It should be noted that Equation~\eqref{eq:voronov_rate_coeff} and the tabulated coefficients assume the Maxwell-Boltzmann (M-B) distribution. The assumption of a M-B EED is a strong one, but one that is commonly used and mathematically well-defined. Since there is no knowledge of the energy distribution of the confined electrons~\cite{izotov18}, we have elected to use the rate coefficient formula based on the assumed M-B EED.

\begin{table}
\centering
\caption{The coefficients for the Voronov semi-empirical analytical equation for the electron-impact ionisation rate coefficient in the case of potassium. $\delta E$ is the ionisation threshold for the charge state $q$ ion, while $P$, $A$, $X$, and $K$ are experimentally determined fitting coefficients. Data corresponding to ionisation from closed electronic shells are highlighted. Data from Ref.~\cite{voronov97}.}
\label{tab:voronov_coefficients}
\begin{tabular}{cccccc}
\hline\hline
$q$ & $\delta E$ (eV) & $P$ & $A$& $X$   & $K$    \Tstrut\Bstrut \\ \hline  
0+     & 4.3    & 1 & 2.02E-07 & 0.272 & 0.31 \Tstrut \\
\rowcolor{pink}
1+     & 31.6   & 1 & 4.01E-08 & 0.371 & 0.22 \\
2+     & 45.8   & 1 & 1.5E-08  & 0.433 & 0.21 \\
3+     & 60.9   & 1 & 1.94E-08 & 0.889 & 0.16 \\
4+     & 82.7   & 1 & 6.95E-09 & 0.494 & 0.18 \\
5+     & 99.4   & 1 & 4.11E-09 & 0.54  & 0.17 \\
6+     & 117.6  & 1 & 2.23E-09 & 0.519 & 0.16 \\
7+     & 154.7  & 1 & 2.15E-09 & 0.828 & 0.14 \\
8+     & 175.8  & 0 & 1.61E-09 & 0.642 & 0.13 \\
\rowcolor{pink}
9+     & 504    & 1 & 1.07E-09 & 0.695 & 0.13 \\
10+    & 564.7  & 1 & 3.78E-10 & 0.173 & 0.3  \\
11+    & 629.4  & 0 & 6.24E-10 & 0.418 & 0.33 \\
12+    & 714.6  & 1 & 2.29E-10 & 0.245 & 0.28 \\
13+    & 786.6  & 1 & 1.86E-10 & 0.344 & 0.23 \\
14+    & 861.1  & 0 & 2.69E-10 & 0.396 & 0.37 \\
15+    & 968    & 1 & 1.06E-10 & 0.912 & 0.13 \\
16+    & 1053.4 & 1 & 4.24E-11 & 0.737 & 0.16 \\
\rowcolor{pink}
17+    & 4610.9 & 0 & 1.38E-11 & 0.416 & 0.34 \\
18+    & 4934.1 & 1 & 3.67E-12 & 0.555 & 0.18 \\ \hline\hline
\end{tabular}
\end{table}

The cross section for charge exchange between an ion at charge state $q$ and a neutral atom can be estimated as~\cite{kronholm19, knudsen81}
\begin{equation}\label{eq:cross_section_cx}
    \sigma^{\mathrm{cx}}_{q\to q-1} =
    \pi r_0^2 q \left(\frac{I_0}{I}\right)^2 Z_{\mathrm{eff},}
\end{equation}
where $r_0$ is the Bohr radius, $I_0$ the ionisation potential of hydrogen, $I$ the ionisation potential of the neutral atom, and $Z_{\mathrm{eff}}$~\cite{slater30, clementi63} its effective proton number as seen by the valence electron after the screening by inner shell electrons has been accounted for. Equation~\eqref{eq:cross_section_cx} is a geometrical cross section based on the Bohr atomic model. 

The $\sigma^\mathrm{cx}_{q\to q-1}$ is independent of the interaction energy up to around 10~keV/u~\cite{knudsen81}, which is vastly in excess of the energy that an ion may realistically gain in the ECRIS plasma ($T_i\sim 10$~eV~\cite{kronholm19, kronholm20}). Therefore, the rate coefficient for charge exchange may be obtained relatively simply as
\begin{equation}\label{eq:rate_coeff_cx}
    \left\langle\sigma v\right\rangle^{\mathrm{cx}}_{q\to q-1} = 
    \sigma^\mathrm{cx}_{q\to q-1}\left\langle v_i^q \right\rangle = 
    \sigma^\mathrm{cx}_{q\to q-1}\sqrt{\frac{8 T_i^q}{\pi m_i}},
\end{equation}

where $\left\langle v_i^q \right\rangle$ is the average speed of ions given a Maxwell-Boltzmann distribution, and $m_i$ is the ion mass. In the above derivation neutrals were assumed cold relative to the ions.

\subsection{Conversion of the balance equation to beam current formalism}\label{sec:conversion}
The extracted beam current for charge state $q$ can be expressed as~\cite{west82, douysset00}
\begin{equation}\label{eq:current_by_west}
    I^q = \kappa \frac{(2L)S}{2}\frac{n^q q e}{\tau^q} \propto \frac{ n^q q}{\tau^q},
\end{equation}
where $\kappa$ is the transmission efficiency of the low energy beamline, $2L$ is the length of the plasma chamber, $S$ the area of the extraction aperture, $n^q$ the number density of ions at charge state $q$, $e$ is the elementary charge, and $\tau^q$ the confinement time of the population of ions at charge state $q$. Assuming, that $\kappa$, $S$ and $L$ are the same for the consecutive charge states $q-1$, $q$ and $q+1$, they can be absorbed into one constant which disappears upon substitution to the balance equation. The balance equation becomes in the extraction current formalism
\begin{equation}\label{eq:balance_eqn_current}
    \begin{split}
        \frac{\mathrm{d}}{\mathrm{d}t}I^q = 
    &n_e \left\langle \sigma v \right\rangle^{\mathrm{inz}}_{q-1 \to q}\frac{q}{q-1}\frac{\tau^{q-1}}{\tau^q}I^{q-1} \\
    - &\left( n_e \left\langle \sigma v \right\rangle^{\mathrm{inz}}_{q \to q+1} + n_0 \left\langle \sigma v \right\rangle^{\mathrm{cx}}_{q \to q-1} + 1/\tau^q \right)I^q \\
    + &n_0\left\langle \sigma v \right\rangle^{\mathrm{cx}}_{q+1 \to q}\frac{q}{q+1}\frac{\tau^{q+1}}{\tau^q}I^{q+1}.
    \end{split}
\end{equation}
By defining:
\begin{align}
    &a_q = n_e \left\langle \sigma v \right\rangle^{\mathrm{inz}}_{q-1 \to q}\frac{q}{q-1}\frac{\tau^{q-1}}{\tau^q} 
    \label{eq:a} \\
    &b_q = \left( n_e \left\langle \sigma v \right\rangle^{\mathrm{inz}}_{q \to q+1} + n_0 \left\langle \sigma v \right\rangle^{\mathrm{cx}}_{q \to q-1} + 1/\tau^q \right) 
    \label{eq:b} \\
    &c_q = n_0\left\langle \sigma v \right\rangle^{\mathrm{cx}}_{q+1 \to q}\frac{q}{q+1}\frac{\tau^{q+1}}{\tau^q},
    \label{eq:c}
\end{align}
equation \eqref{eq:balance_eqn_current} becomes
\begin{equation}\label{eq:fitting_function_ddt}
    \frac{\mathrm{d}}{\mathrm{d}t}I^q = a_q I^{q-1} - b_q I^q + c_q I^{q+1}.
\end{equation}
Equation~\eqref{eq:fitting_function_ddt} may be used to determine the parameters $a_q$, $b_q$, and $c_q$ by fitting to experimentally measured beam current transients $\mathrm{d}I^q/\mathrm{d}t$. This procedure is described in section~\ref{sec:determining_coefficients}.

\subsection{Deconvolution of the characteristic values from the fitting parameters}\label{sec:determ plasma param}

The definitions \eqref{eq:a}, \eqref{eq:b} and \eqref{eq:c} hold for all $q$. From equation \eqref{eq:c} we obtain (by choosing $q\to q-1$): 
\begin{equation}
    c_{q-1} = n_0\left\langle \sigma v \right\rangle^{\mathrm{cx}}_{q \to q-1}\frac{q-1}{q}\frac{\tau^{q}}{\tau^{q-1}},
\end{equation}
which can be rearranged to obtain
\begin{equation}\label{eq:intermediate_1}
    n_0\left\langle \sigma v \right\rangle^{\mathrm{cx}}_{q \to q-1} =
    \frac{q}{q-1}\frac{\tau^{q-1}}{\tau^q}c_{q-1}.
\end{equation}
The fraction $\tau^{q-1}/\tau^q$ can be obtained from equation~\eqref{eq:a} as 
\begin{equation}\label{eq:intermediate_2}
    \frac{\tau^{q-1}}{\tau^q} = \frac{q-1}{q}\frac{a_q}{n_e\left\langle\sigma v\right\rangle^{\mathrm{inz}}_{q-1\to q}}.
\end{equation}
By substituting equations~\eqref{eq:intermediate_1} and~\eqref{eq:intermediate_2} into equation~\eqref{eq:b}, we obtain an expression for the confinement time:
\begin{equation}\label{eq:tau}
    \tau^q = \left( b_q 
    - n_e \left\langle\sigma v\right\rangle^{\mathrm{inz}}_{q\to q+1}
    - \frac{a_q c_{q-1}}{n_e\left\langle\sigma v\right\rangle^{\mathrm{inz}}_{q-1\to q}}
    \right)^{-1},
\end{equation}
which is a function of $n_e, T_e$ and the fitting parameters. From equation~\eqref{eq:intermediate_2} (making the substitution $q \to q+1$) one obtains
\begin{equation}\label{eq:tau_from_a}
    \frac{q}{q+1}\frac{ a_{q+1}}{n_e\left\langle\sigma v\right\rangle^{\mathrm{inz}}_{q\to q+1}}  = \frac{\tau^q}{\tau^{q+1}}
    .
\end{equation}
The confinement times in Equation~\eqref{eq:tau_from_a} can be calculated using Equation~\eqref{eq:tau} (simply substituting $q\to q+1$ to obtain $\tau^{q+1}$). This leads to the following equation for $n_e$ and $T_e$:
\begin{equation}\label{eq:eqn_of_ne_Te}
    \frac{q}{q+1}\frac{ a_{q+1}}{n_e\left\langle\sigma v\right\rangle^{\mathrm{inz}}_{q\to q+1}}  =
     \frac
     {b_{q+1} - n_e \left\langle\sigma v\right\rangle^{\mathrm{inz}}_{q+1\to q+2}
    - \frac{a_{q+1} c_{q}}{n_e\left\langle\sigma v\right\rangle^{\mathrm{inz}}_{q\to q+1}}}
     {b_q - n_e \left\langle\sigma v\right\rangle^{\mathrm{inz}}_{q\to q+1}
    - \frac{a_q c_{q-1}}{n_e\left\langle\sigma v\right\rangle^{\mathrm{inz}}_{q-1\to q}}}
\end{equation}
where the $T_e$ dependence resides in the ionisation rate coefficients. Having determined the coefficients $a_q, b_q,$ and $c_q$ from the experimental data it is possible to search for the allowed pairs of the parameters $n_e$ and $T_e$ which satisfy Equation~\eqref{eq:eqn_of_ne_Te}. Mathematically the equation of two unknowns has an infinitude of solutions, but the solution space may be constrained by physical considerations as explained in Section~\ref{sec:numerical_method}. The solution pairs constitute the set of \textit{postdictions} for the possible range of $(n_e, T_e)$ values in the support plasma, which could account for the measured $a_q, b_q, c_q$ coefficients.

The ionisation time ($1/n_e\left\langle\sigma v\right\rangle^{\text{inz}}_{q\to q+1}$) can be calculated using the obtained $(n_e, T_e)$-pairs and Equation~\eqref{eq:voronov_rate_coeff}. One may then also calculate from the definition~\eqref{eq:b} of the parameter $b_q$, the characteristic charge exchange time:
\begin{equation}\label{eq:cx_time}
    1/n_0\left\langle\sigma v\right\rangle^{\text{cx}}_{q\to q-1} =
    \left(b_q - 1/n_e\left\langle\sigma v\right\rangle^{\text{inz}}_{q\to q+1} - 1/\tau^q\right)^{-1}
\end{equation}

Note, that formulas~\eqref{eq:tau}, \eqref{eq:tau_from_a}, \eqref{eq:eqn_of_ne_Te}, and \eqref{eq:cx_time} do not describe physical dependencies of observables on plasma parameters, as they are merely formulas for calculating their values based on the fitting coefficients, and other plasma parameters.

\section{Numerical methods}\label{sec:numerical_method}
\subsection{Determining coefficients $a_q, b_q$ and $c_q$}\label{sec:determining_coefficients}
Equation~\eqref{eq:fitting_function_ddt} can be used to obtain the coefficients $a_q, b_q, c_q$ by fitting to the measurement data, but due to the noise involved in the current measurement, calculating the derivative to make the fits is not practical. We instead used the 4$^\text{th}$ order Runge-Kutta method to solve for $\mathfrak{I}^q(t)$ the system
\begin{equation}\label{eq:system for RK4}
    \begin{cases}
        I^{q-1}(t),\\ 
        \dot {\mathfrak{I}}^q (t) = l I^{q-1}(t) - m\mathfrak{I}^q(t) + n I^{q+1}(t), \\ 
        I^{q+1}(t),
    \end{cases}
\end{equation}
where $I^{q-1}$ and $I^{q+1}$ are taken from the measurement, and coefficients $l,m,n$ can be varied. The residual of the measured current $I^q(t)$, and $\mathfrak{I}^q(t)$ is then minimised by the least squares method where the $l, m$ and $n$ corresponding to the minimum residue are taken to represent $a_q$, $b_q$, and $c_q$. 

We have assumed here, that the plasma parameters are constants in time, i.e. that the perturbation on the support plasma caused by the 1+ injection pulse is sufficiently small, for $n_e$, $T_e$, $n_0$, the ion temperatures $T_i^q$ and confinement times $\tau^q$ to be determined by the support plasma. The validity of this assumption is discussed in section~\ref{sec:discussion}.

\subsection{Determining the plasma parameters}\label{sec:determining_energy_content}
In order to find the pairs $n_e, T_e$, which satisfy the Equation~\eqref{eq:eqn_of_ne_Te}, we defined a penalty function, namely the deviation of the two terms set by Eq.~\eqref{eq:eqn_of_ne_Te}
\begin{equation}
    \begin{split}
        &F^q(n_e, T_e) \equiv \\
        &\left\vert 
        \frac{q}{q+1}\frac{ a_{q+1}}{n_e\left\langle\sigma v\right\rangle^{\mathrm{inz}}_{q\to q+1}} -
         \frac
         {b_{q+1} - n_e \left\langle\sigma v\right\rangle^{\mathrm{inz}}_{q+1\to q+2}
        - \frac{a_{q+1} c_{q}}{n_e\left\langle\sigma v\right\rangle^{\mathrm{inz}}_{q\to q+1}}}
         {b_q - n_e \left\langle\sigma v\right\rangle^{\mathrm{inz}}_{q\to q+1}
        - \frac{a_q c_{q-1}}{n_e\left\langle\sigma v\right\rangle^{\mathrm{inz}}_{q-1\to q}}}
        \right\vert
        \bigg / \frac{q}{q+1}\frac{ a_{q+1}}{n_e\left\langle\sigma v\right\rangle^{\mathrm{inz}}_{q\to q+1}}
    \end{split}
\end{equation}
which was minimised with the constraints
\begin{align}
    \begin{cases}
    0 < \tau^{q}, \\
    0 < \tau^{q+1},\\
    0 <  1/n_0\left\langle\sigma v\right\rangle^{\text{cx}}_{q\to q-1}, \\ 
    n_{e,\text{low}} < n_e < n_{e,\text{co}},\\
    T_{e,\text{low}} < T_e < T_{e,\text{high}}
    \end{cases}
\end{align}
where $\tau^q$ and $\tau^{q+1}$ were calculated using Equation~\eqref{eq:tau}, and $ 1/n_0\left\langle\sigma v\right\rangle^{\text{cx}}_{q\to q-1}$ using Equation~\eqref{eq:cx_time}. The first three constraints are intuitive as they merely state that the characteristic times are positive. The upper limit for the support plasma electron density $n_e$ can be taken to be the cut-off density which can be calculated from~\cite{geller_ecris}
\begin{equation}
    n_{e,\text{co}} = \frac{\varepsilon_0 m_e(2\pi f)^2}{e^2},
\end{equation}
where $\varepsilon_0$ is the vacuum permittivity, $m_e$ is electron mass, $e$ the elementary charge and $f$ is the frequency of the inbound microwave. For the 14.5~GHz microwaves this value is $2.61\times10^{12}$~cm$^{-3}$. The lower limit for $n_e$ can be approximated based on the results from Ref.~\cite{tarvainen16}, where it was found that the lower limit for $n_e$ increases as a function of the microwave power: Using 14.5~GHz frequency and 470~W microwave power they found $n_{e,\text{low}}$ to be $4.4\times10^{11}$~cm$^{-3}$. As a conservative lower limit, we take  $n_{e,\text{low}} = 1\times10^{11}$~cm$^{-3}$ for the 500~W microwave power used herein. The lower limit for the electron temperature is positive and non-zero, and $T_{e,\text{low}} = 10$~eV was chosen as it corresponds to the order of magnitude of the plasma potential~\cite{lamy11}. The upper limit $T_{e,\text{high}}$ is set to 10~keV, as according to Ref.~\cite{perret98} (as cited in Ref.~\cite{douysset00}) the fraction of electrons having an energy higher than a few keV is between 20~\% and 50~\%. This corresponds to the warm electron population, which mostly accounts for the ionisation processes within the plasma. Based on an analysis of the solution sets (see supplement~\ref{supplement:penalty_function}) only such minima for which $F^q(n_e,T_e) < 10^{-4}$ were accepted, and solutions leading to negative (i.e. unphysical) results were discarded.

The semi-empirical analytical expression~\eqref{eq:voronov_rate_coeff} was used to evaluate the necessary rate coefficients. In order to account for the uncertainty of Eq.~\eqref{eq:voronov_rate_coeff}, a Monte Carlo approach was employed, where a random systematic bias --- within the uncertainty limits given for the formula --- was added to each rate coefficient involved in the function $F^q(n_e,T_e)$. According to Voronov in Ref.~\cite{voronov97}, the uncertainty of the rate coefficients is $40~\%-60~\%$.

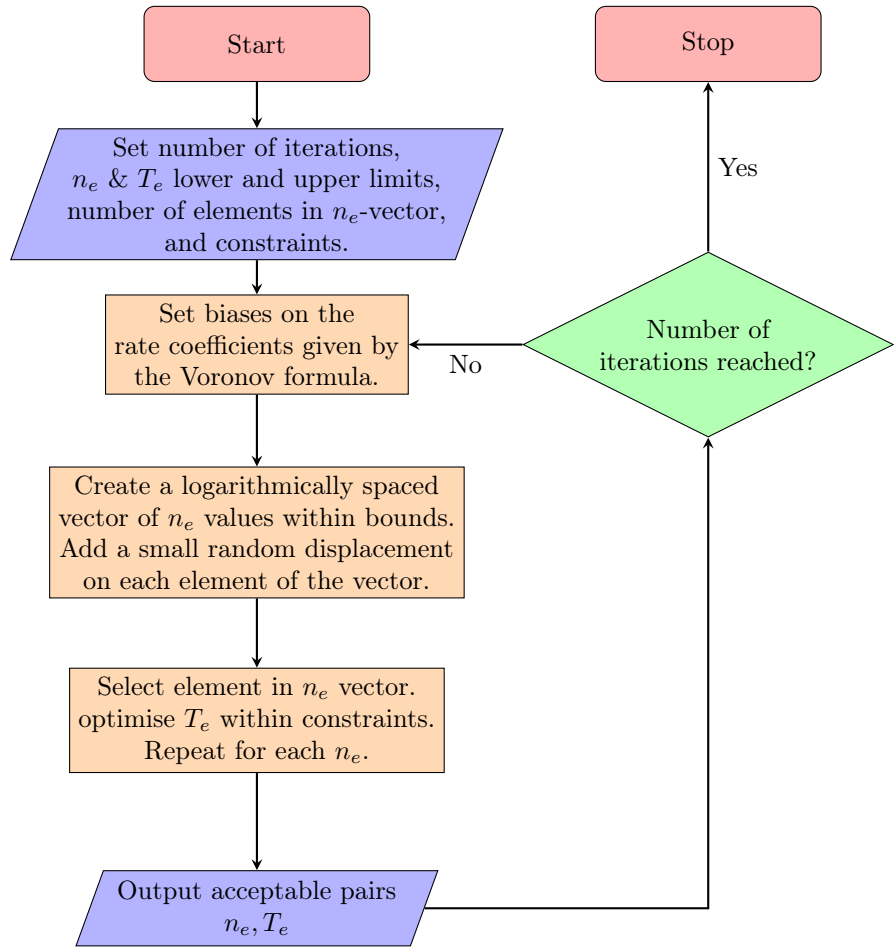
\begin{figure}
    \centering
        \begin{tikzpicture}[node distance = 2 cm, scale = 0.9]
        
        \node 
        (start) 
        [startstop]
        {Start};
        
        \node
        (input)
        [io, below of = start, draw, align = center]
        {Set number of iterations,\\
        $n_e$ \& $T_e$ lower and upper limits,\\
        number of elements in $n_e$-vector,\\
        and constraints.};

        \node 
        (mc)
        [process, below of = input, draw, align = center]
        {Set biases on the \\
        rate coefficients given by \\
        the Voronov formula.};
        
        \node
        (pro1)
        [process, below of = mc, draw, align = center, 
        yshift = -0.5 cm]
        {Create a logarithmically spaced \\
        vector of $n_e$ values within bounds. \\ 
        Add a small random displacement\\
        on each element of the vector.};
        
        \node
        (pro2)
        [process, below of = pro1, draw, align = center,
        yshift = -0.5 cm]
        {Select element in $n_e$ vector.\\
        optimise $T_e$ within constraints.\\
        Repeat for each $n_e$.};
        
        \node
        (decision1)
        [decision, right of = mc, draw, align = center,
        aspect=2, xshift = 4 cm, yshift = 0 cm]
        {Number of\\ iterations reached?};

        \node
        (out)
        [io, below of = pro2, draw, align = center,
        yshift = -0.5 cm]
        {Output acceptable pairs \\ $n_e, T_e$};
        
        \node
        (stop)
        [startstop, right of = start, 
        xshift= +4 cm]
        {Stop};
        
        \draw [arrow] (start) -- (input);
        \draw [arrow] (input) -- (mc);
        \draw [arrow] (mc) -- (pro1);
        \draw [arrow] (pro1) -- (pro2);
        \draw [arrow] (pro2) -- (out);
        \draw [arrow] (out) -| (decision1);
        \draw [arrow]  (decision1) -- node[anchor=west]{Yes}(stop);
        \draw [arrow]  (decision1) -- node[anchor=north]{No}(mc);
        
        \end{tikzpicture}
    \caption{Flowchart of the code designed to search for the acceptable $(n_e, T_e)$-pairs satisfying Equation~\eqref{eq:eqn_of_ne_Te}.}
    \label{fig:flowchart_poptau}
\end{figure}

The flow of the code is illustrated in Figure~\ref{fig:flowchart_poptau}. The algorithm was run using 1000 different, randomized uncertainty biases to ensure that the solution set was sufficiently sampled. The process produces a set of $(n_e, T_e)$-pairs, which minimise the penalty function within the given constraints. When the confinement times, ionisation times, charge exchange times and plasma energy contents are calculated for each pair, we obtain a distribution of results for each quantity, respectively. We take the median, and the 34.1-percentile below and above the median to represent the results, which corresponds to a one sigma uncertainty for a normal distribution.

The $N+$ pulse response of three consecutive charge states ($I^{q-1}$, $I^q$, and $I^{q+1}$) are necessary to estimate the $a_q$, $b_q$, and $c_q$ for any given charge state. To solve Eq.~\eqref{eq:eqn_of_ne_Te}, on the other hand, requires the parameters $a_q$, $a_{q+1}$, $b_q$, $b_{q+1}$, $c_{q-1}$, and $c_q$. To obtain them, five neighboring charge state currents must be measured. By extension, given $X$ measured neighboring currents in the CSD, one can obtain a set of solutions for $X-4$ charge states.

\section{Experimental methods}\label{sec:experimental_methods}
The method consists of injecting short pulses of 1+ ions into a CB-ECRIS plasma and analysing the extracted n+ responses. High 1+ capture efficiencies, ranging between 50~\% and 60~$\%$, are typically measured for injected beam intensities up to 1~$\mu$A in continuous 1+ injection mode (CW mode) \cite{angot_2012,lamy_proc_ecris_2014}. The optimum 1+ ion capture efficiency is obtained when the velocity of the injected ions is equal to the average velocity of the support plasma ions \cite{delcroix94}. It can be finely tuned by adjusting the potential difference $\Delta$V between the source generating the 1+ beam and the charge breeder. The number of injected particles is managed by tuning the 1+ source to produce the chosen 1+ beam intensity in order to minimise the perturbation of the buffer plasma.

\subsection{Experimental setup}\label{sec:experimental_setup}
The measurements were conducted by injecting a K$^+$ beam into a He support plasma to obtain high charge breeding efficiencies~\cite{maunoury_ecris_2016}. K was chosen as the injected species because it is an alkali, so there is no recycling of the ions lost on the charge breeder plasma chamber wall into the plasma, and several consecutive charge states, ranging from 1+ to 12+, can be measured. It is also the heaviest alkali for which cross section / rate coefficient data was available.

\begin{figure}
    \centering
    \includegraphics[width=1.0\textwidth]{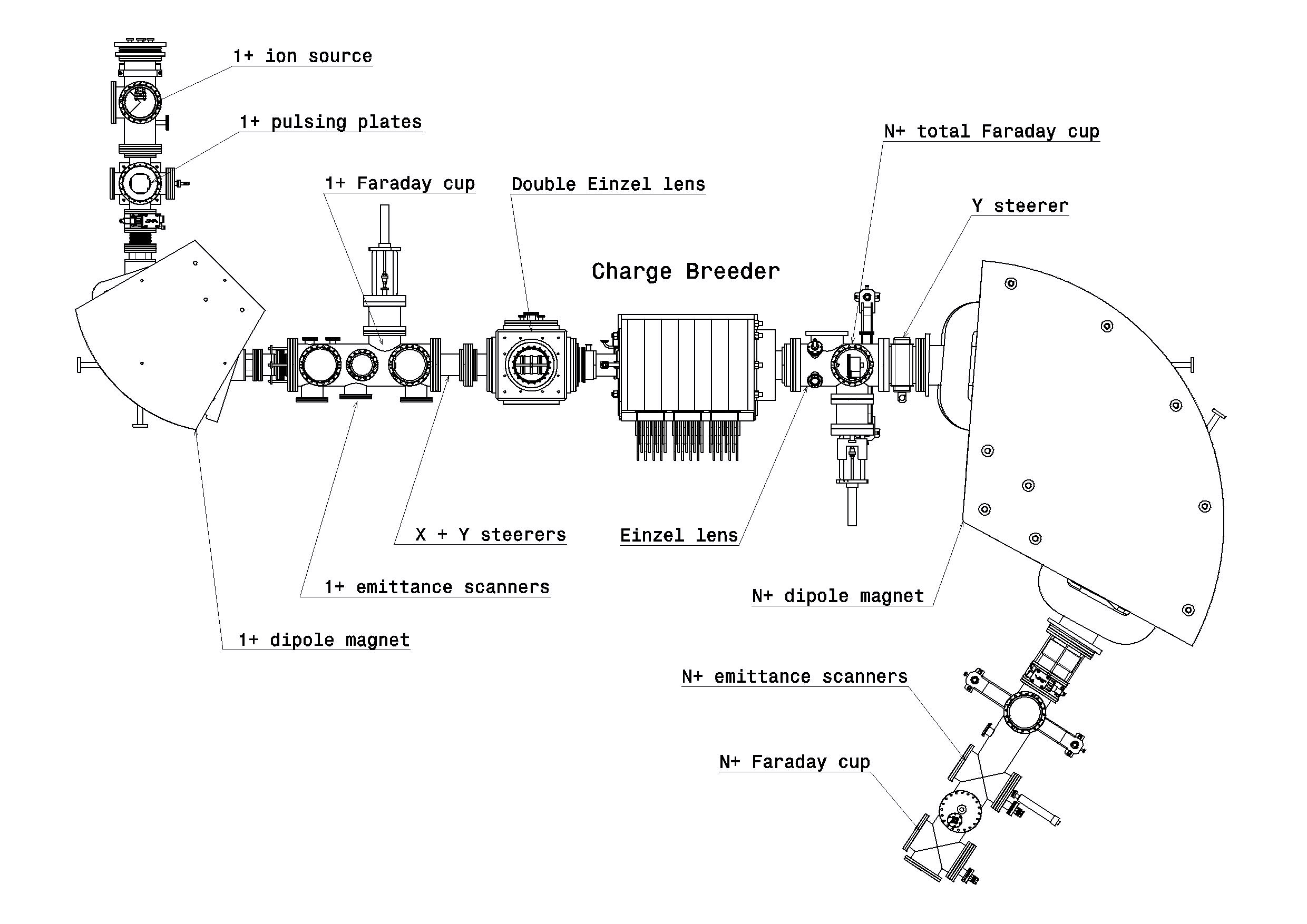}
    \caption{schematic layout of the 1+ $\to$ N+ test bench.}
    \label{fig:1+N+beamline}
\end{figure}

Experiments were conducted on an upgraded version of the Laboratory of Subatomic Physics \& Cosmology (LPSC) 1+$\rightarrow$N+ test bench~\cite{angot_2020}, see Figure~\ref{fig:1+N+beamline}, with respect to the configuration for the short pulse injection studies done previously \cite{angot18}. After these modifications, which essentially consisted of improving the vacuum, and surface residue mitigation; the background vacuum pressure at injection was $2.5 \times 10^{-8}$~mbar.

The charge breeder was assembled with a hexapole providing a 0.8~T radial magnetic field strength at plasma chamber wall, on the poles. An additional soft iron plug was mounted under vacuum to increase the axial magnetic field strength at injection \cite{angot_icis_2017}.
The plasma electrode aperture diameter was 8~mm. For the experiments, the charge breeder was operated at 20~kV extraction voltage with a He plasma, the extracted beams being mass-analysed using the N+ dipole, and measured at the N+ Faraday Cup (FC) (see Fig.~\ref{fig:1+N+beamline}).

First, the charge breeder was tuned to optimise the K$^{10+}$ efficiency in continuous 1+ injection mode. A 0.71 $\mu$A ($0.44\times10^{13}$ pps) K$^+$ beam was produced with the ion gun 1+ source   \cite{lamy_proc_ecris_2014}. The K$^+$ beam was selected by the 1+ beam line dipole magnet and injected into the charge breeder.  The electrostatic pulsing system, set just before the 1+ dipole magnet, was used to pulse the 1+ beam into the charge breeder in order to calculate the charge breeding efficiencies taking into account the N+ background.

The $\Delta V$ parameter was carefully adjusted to optimise the capture of the injected 1+ ions. The optimum K$^{10+}$ efficiency was found with the charge breeder axial magnetic field strength values of $B_{\mathrm{inj}} =1.58$~T, $B_{\mathrm{min}} = 0.45$~T and $B_{\mathrm{ext}} = 0.83$~T, the 14.5~GHz microwave power being set at 500~W. The vacuum level at injection was  $8.7 \times 10^{-8}$~mbar and the $\Delta V$  was set at -3.9~V.  In this configuration, the ECR zone length on the charge breeder axis was simulated to be 122~mm.

Table~\ref{tab:measured_efficiencies} summarizes the measured charge breeding efficiencies of charge states between K$^{+}$ and K$^{12+}$. For reference, the best efficiency measured for K with He as a support gas was 11.7~$\%$ for K$^{10+}$ in the previous test bench configuration~\cite{angot_icis_2017}, i.e. the data were taken in conditions relevant for the charge breeding process.
\begin{table}
\setlength{\tabcolsep}{4.5pt} %% default is 6pt
\centering
    \caption{Measured charge breeding efficiencies for charge states between K$^{+}$ and K$^{12+}$.}
    \label{tab:measured_efficiencies}
      \begin{tabular}{ccccccccccccc}
        \hline\hline\Tstrut\Bstrut
            Ion & K$^+$ & K$^{2+}$ & K$^{3+}$ & K$^{4+}$ & K$^{5+}$ & K$^{6+}$ & K$^{7+}$ & K$^{8+}$ & K$^{9+}$ & K$^{10+}$ & K$^{11+}$ & K$^{12+}$\\
            \midrule
            Efficiency ($\%$) & 9.7 & 2.7 & 1.2 & 1.0 & 1.1 & 1.2 & 1.5 & 2.7 & 8.9 & 10.6 & 8.5 & 5.1\\
        \hline\hline\Tstrut\Bstrut
    \end{tabular}
\end{table}

The influence of the 1+ beam injection on the plasma was checked in continuous mode, comparing the CSD extracted from the CB with and without 1+ injection.
Without 1+ injection, the He$^{+}$ and He$^{2+}$  beam intensities were 65.2~$\mu$A and 19.6~$\mu$A, respectively. The measured variations of these peaks when injecting the K$^{+}$ beam were -0.2~$\%$ for He$^+$ and -3.6~$\%$ for He$^{2+}$. It is worth noting that the He$^{+}$ peak is superimposed with the O$^{4+}$ peak, and He$^{2+}$ peak with H$_{2}^{+}$ peak --- O and H$_2$ being present in the plasma as contaminants. The estimated total flux of He ions extracted from the CB (with the contaminants contribution subtracted) was $36\times10^{13}$ pps. Therefore, in this continuous mode of operation, the K$^{+}$ flux amounts to only about 1~$\%$ of the support gas extracted ions. 

In order to analyse the effect on higher charge states, the change on the oxygen, carbon and nitrogen impurities present in the CSD was also inspected. A slight decrease of the high charge state beam intensities is noticed for O, C and N ion populations with a maximum difference of about -5.4~$\%$ for O$^{7+}$ (see Table~\ref{tab:CSD_change}). This CSD modification is attributed to the gas mixing effect, due to the mass difference between the injected ions and the plasma support gas~\cite{delcroix94,melin99}. Taking into account that in continuous mode $(i)$ the flux of injected K$^{+}$ ions is small compared to the extracted He ions flux (and by extension even smaller compared to the total number of buffer gas ions in the plasma volume) and $(ii)$ the effect on the plasma species is small, we consider here that the 5~ms 1+ beam pulse effect on the support plasma is negligible.
\begin{table}
\centering
    \caption{N+ beam intensities $I$ without 1+ beam injection and N+ beam intensities change $\Delta$ when injecting K$^+$ beam compared to without injection.}
    \label{tab:CSD_change}
    \begin{tabular}{ccccccccc}
        \hline\hline\Tstrut\Bstrut
            Species & He$^{+}$/O$^{4+}$  & He$^{2+}$/H$_{2}^{+}$ & O$^{+}$ & O$^{2+}$ & O$^{3+}$ & O$^{5+}$  & O$^{6+}$  & O$^{7+}$  \\
            \midrule
            $I$~($\mu$A) & 65.2 & 19.6 & 2.9 & 4.1 & 6.5 & 22.9 & 53.1 & 13.4\\
             \midrule           
            $\Delta$~($\%$) & -0.2 & -3.6 & -0.7 & -0.2 & -0.3 &  -2.8 & -3.8 & -5.4\\
           \hline\hline\Tstrut\Bstrut
    \end{tabular}
\end{table}

After the aforementioned measurements, the 1+ injection was switched to pulse mode. Short 1+ pulses with a width of 5~ms, corresponding to $2.2\times10^{10}$ particles per pulse, were injected using a square signal to drive the pulsing system. The 5~ms duration was chosen to obtain N+ pulse responses with a good signal-to-noise ratio, without accumulation effect~\cite{angot18}. The repetition rate was carefully tuned to leave N+ pulse responses enough time to recover between consecutive pulses. The N+ pulse responses were measured with the N+ FC for charge states ranging from K$^{+}$ to K$^{12+}$. The FC was connected to ground via a 5.7~M$\Omega$ resistor. An oscilloscope was used to measure the voltage at the resistor ends and to average 64 times the pulse responses before sending the measurements to the command and control computer for saving. It is worth noting that in the CSD, due to the resolution of the N+ spectrometer, K$^{11+}$ and K$^{12+}$ peaks overlapped with the fringes of N$^{4+}$ and O$^{5+}$, respectively. The change of the O$^{5+}$ and N$^{4+}$ beam intensities in CW mode being -2.8~$\%$ and +1.7~$\%$ compared to the case without 1+ injection, the effect of these contaminants on the K$^{11+}$ and K$^{12+}$ transients was considered negligible in short pulse mode.

\subsection{Measurements and fitting coefficient determination}
\begin{figure}
\begin{subfigure}{0.49\textwidth}
\includegraphics[width=\textwidth]{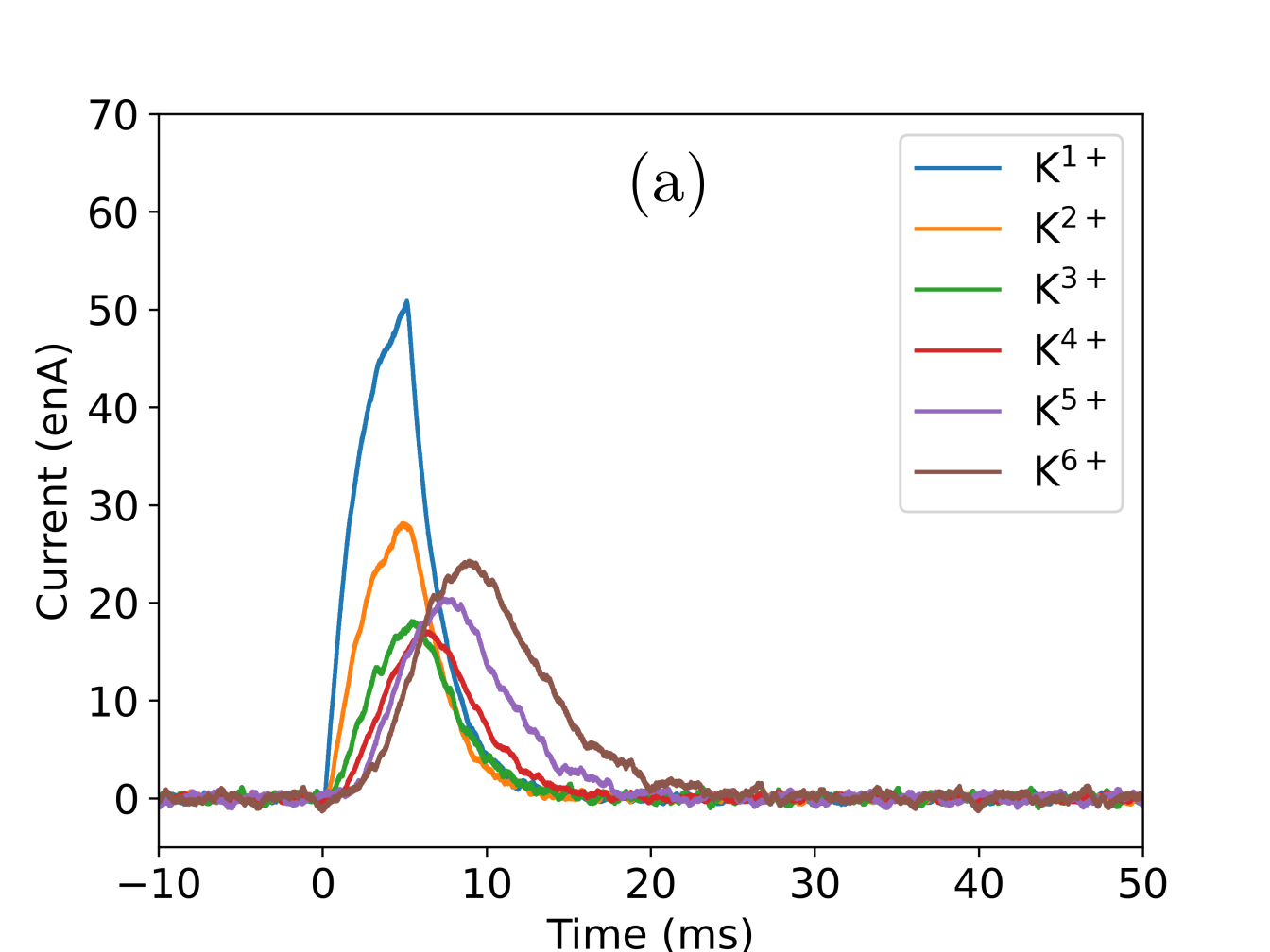} 
\label{fig:subim1}
\end{subfigure}
\begin{subfigure}{0.49\textwidth}
\includegraphics[width=\textwidth]{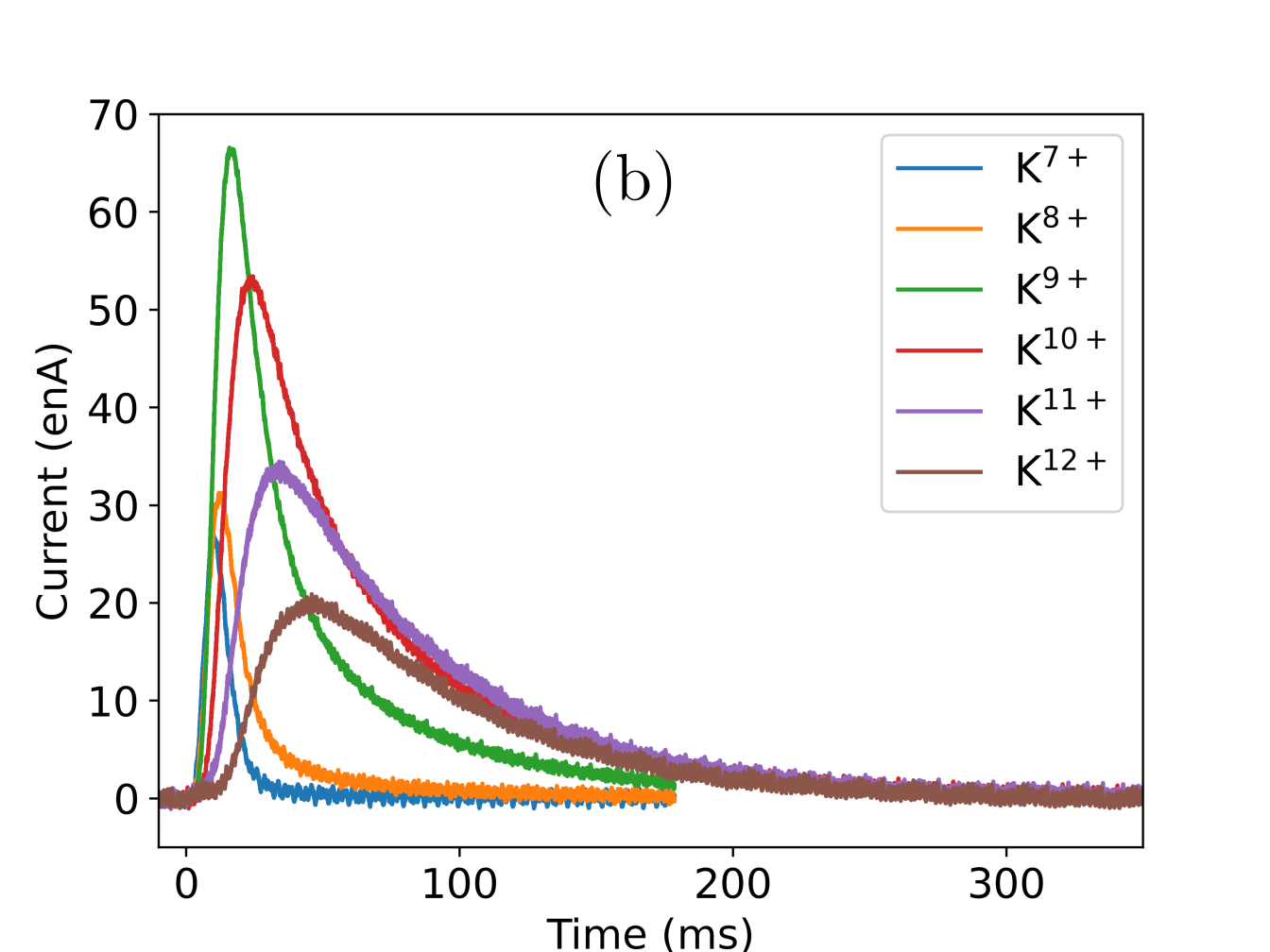}
\label{fig:subim2}
\end{subfigure}
\caption{Extracted K$^{n+}$ pulse responses for charge states between a) 1+ to 6+ and b) 7+ to 12+}
\label{fig:N+ responses}
\end{figure}

Figure~\ref{fig:N+ responses} a) and b) show the N+ pulse responses for charge states from K$^{1+}$ to K$^{12+}$. The onset time of the 1+ injection pulse was set as $t=0$. For each N+ pulse response, the background was calculated by making an average of the response before the pulse start. These background values were subtracted from the respective N+ pulse responses. 

The method described in \ref{sec:determining_coefficients} was applied to estimate the $a_q$, $b_q$ and $c_q$ parameters from the measured currents. The dependence of the fitting coefficients on the fitting range was checked by limiting the fitting window end point. It was found that for K$^{2+}$ and K$^{3+}$ the coefficients did show a noticeable dependence on the fitting range, while for higher charge states such a dependence was not found. This dependence for low charge states is probably caused by in-flight ionisation effects~\cite{tarvainen15} which are not accounted for by the fitting model. Due to the time dependence, only the parameters for charge states K$^{4+}$ and higher were retained. These values are summarized in  Table~\ref{tab:a b c parameters}. More details on the fitting range analysis can be found in the supplementary material in~\ref{supplement:fitting_range}.

\begin{table}[]
\centering
\caption{Calculated  $a_q$, $b_q$ and $c_q$ parameters for K$^{4+}$--K$^{11+}$, with an RK4 stepsize of 10~$\mu$s, and the corresponding reduced $\chi^2$ values of the fits.}
\label{tab:a b c parameters}
\begin{tabular}{ccccccccc}
\hline\hline\Tstrut\Bstrut
     & K$^{4+}$    & K$^{5+}$    & K$^{6+}$    & K$^{7+}$    & K$^{8+}$    & K$^{9+}$    & K$^{10+}$   & K$^{11+}$   \\\hline\Tstrut\Bstrut
$a_q\ (1/\mathrm{s})$    & 931  & 996  & 971  & 855  & 793  & 705  & 175  & 132  \\
$b_q\ (1/\mathrm{s})$    & 894  & 774  & 846  & 840  & 760  & 326  & 231  & 256  \\
$c_q\ (1/\mathrm{s})$    & 2    & 27   & 112  & 136  & 93   & 108  & 117  & 151  \\ \midrule\Tstrut\Bstrut
$\chi^2$ & 1.15 & 1.01 & 1.01 & 1.35 & 1.68 & 2.77 & 1.12 & 1.26 \\ \hline\hline\Tstrut\Bstrut
\end{tabular}
\end{table}

\section{Results}\label{sec:results}
\begin{figure}
\begin{subfigure}{0.48\textwidth}
\includegraphics[width=\linewidth]{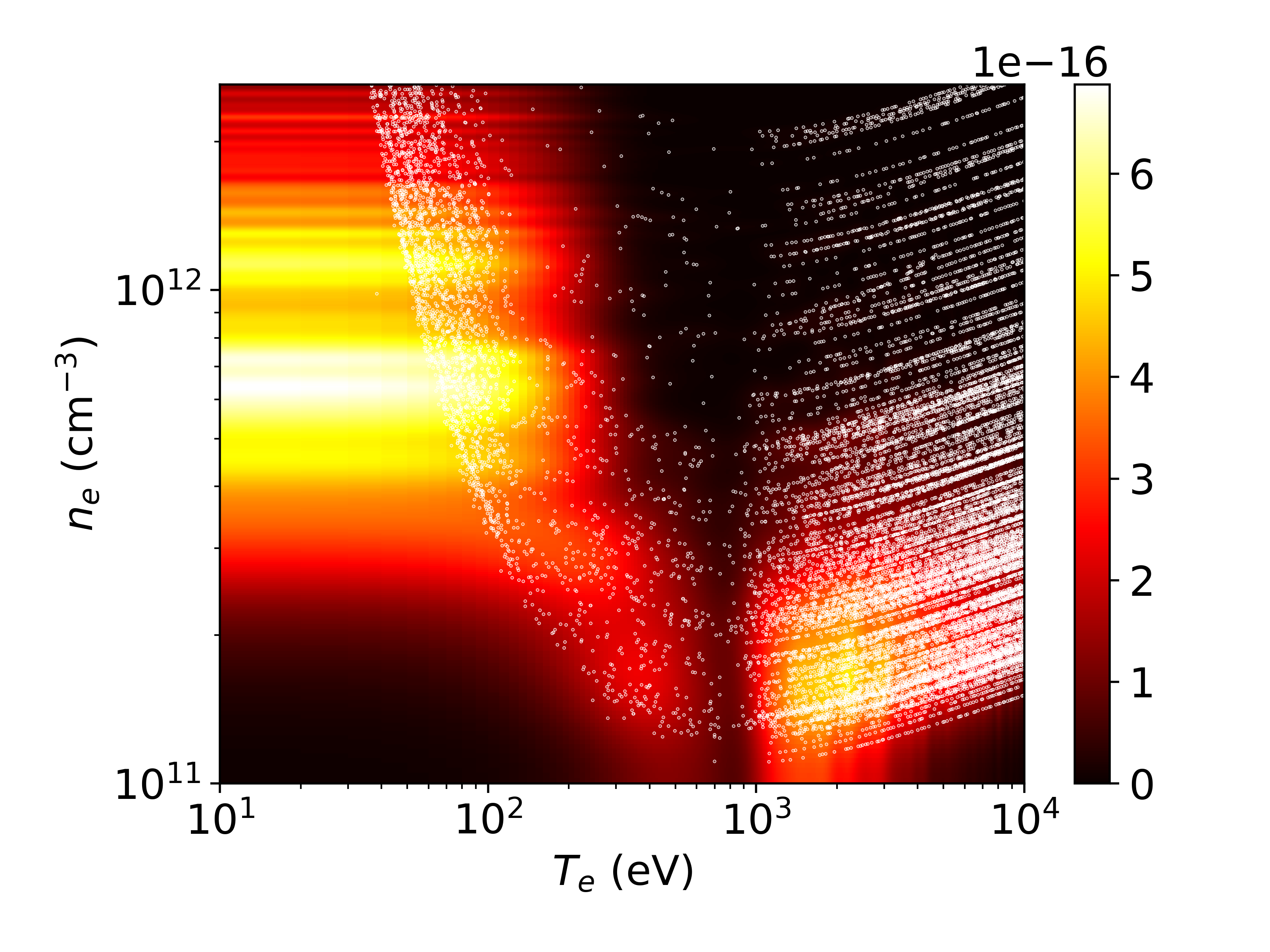}
\caption{K$^{5+}$} \label{fig:a}
\end{subfigure}\hspace*{\fill}
\begin{subfigure}{0.48\textwidth}
\includegraphics[width=\linewidth]{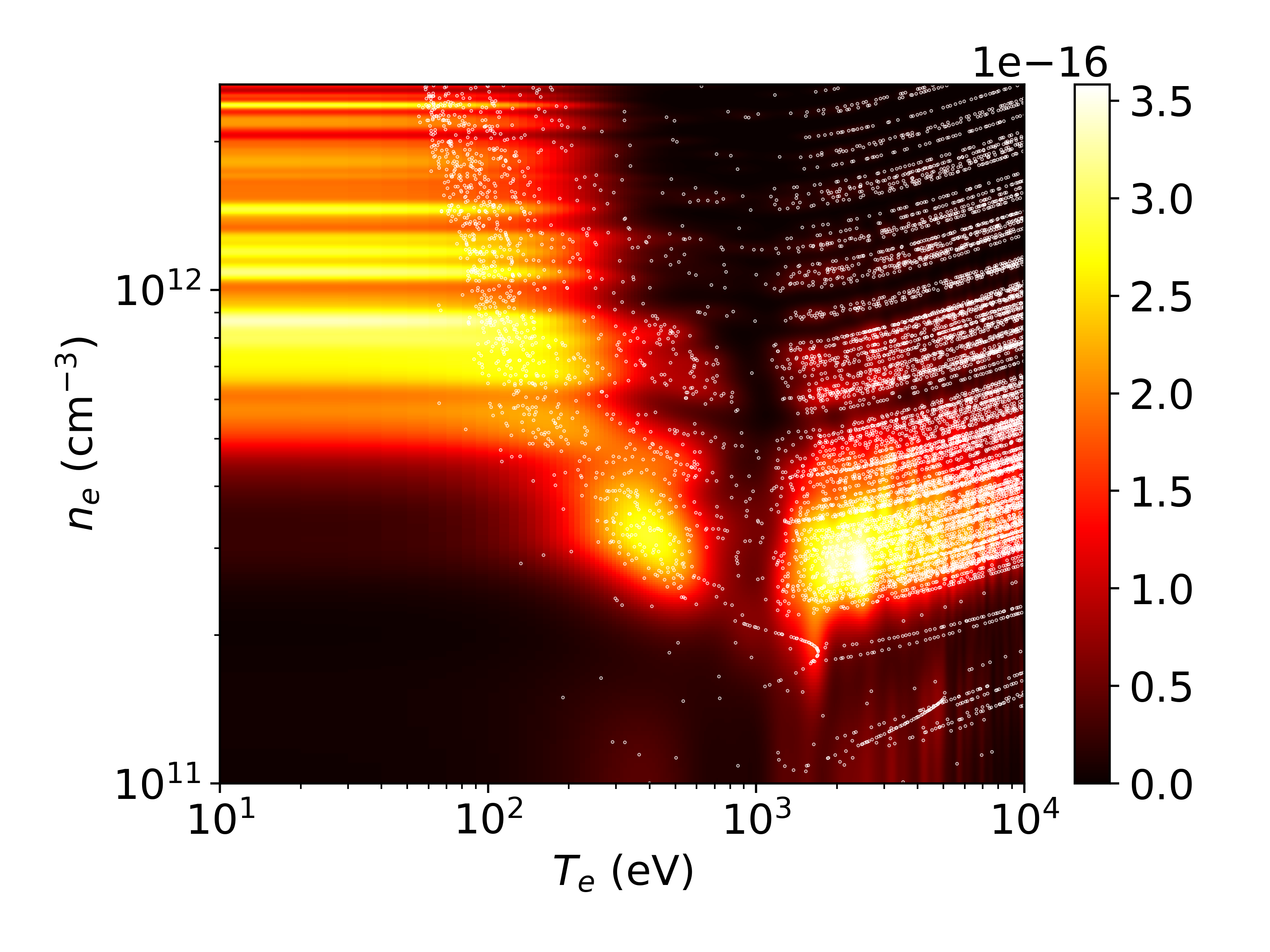}
\caption{K$^{6+}$} \label{fig:b}
\end{subfigure}
\medskip
\begin{subfigure}{0.48\textwidth}
\includegraphics[width=\linewidth]{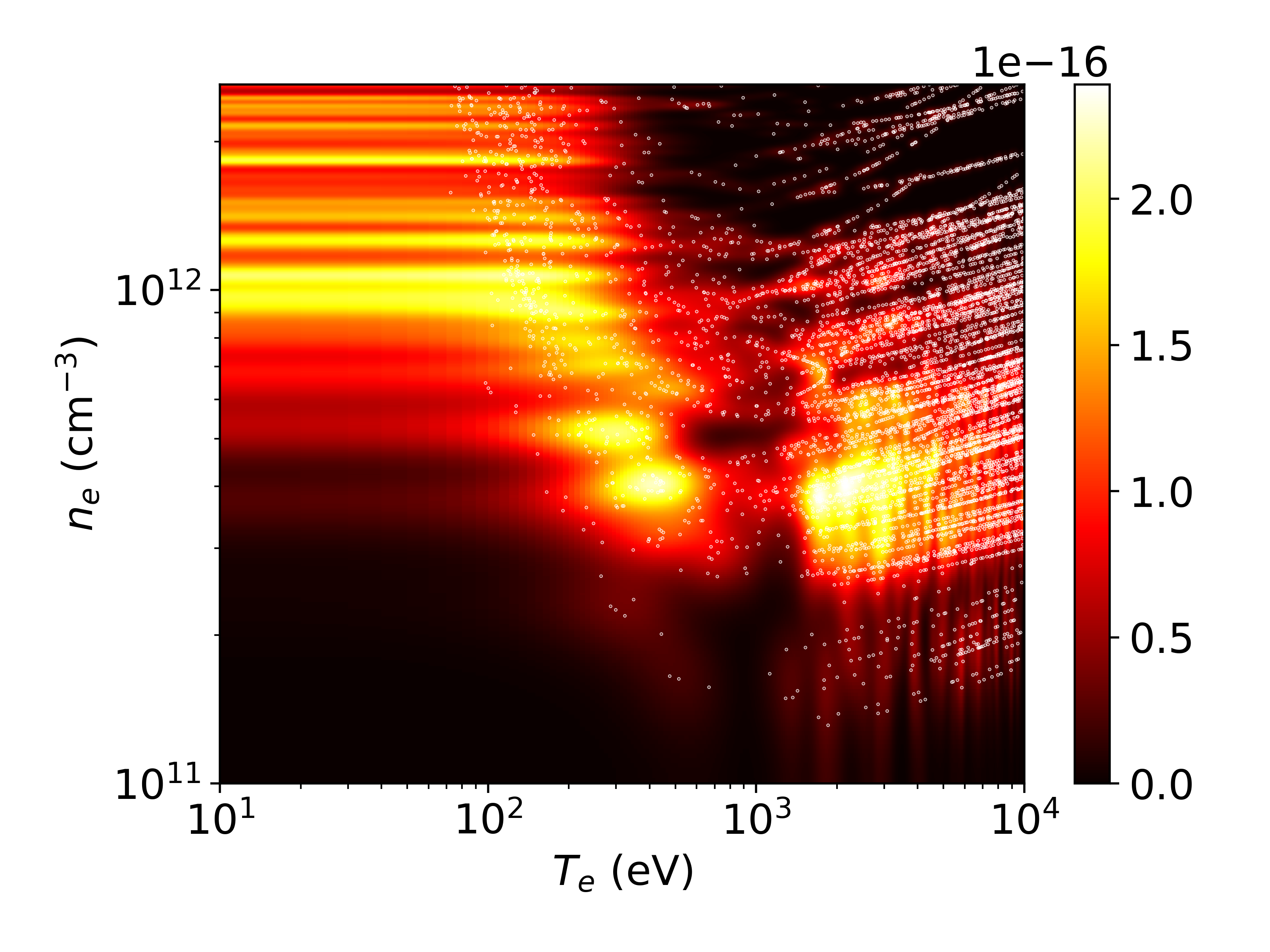}
\caption{K$^{7+}$} \label{fig:c}
\end{subfigure}\hspace*{\fill}
\begin{subfigure}{0.48\textwidth}
\includegraphics[width=\linewidth]{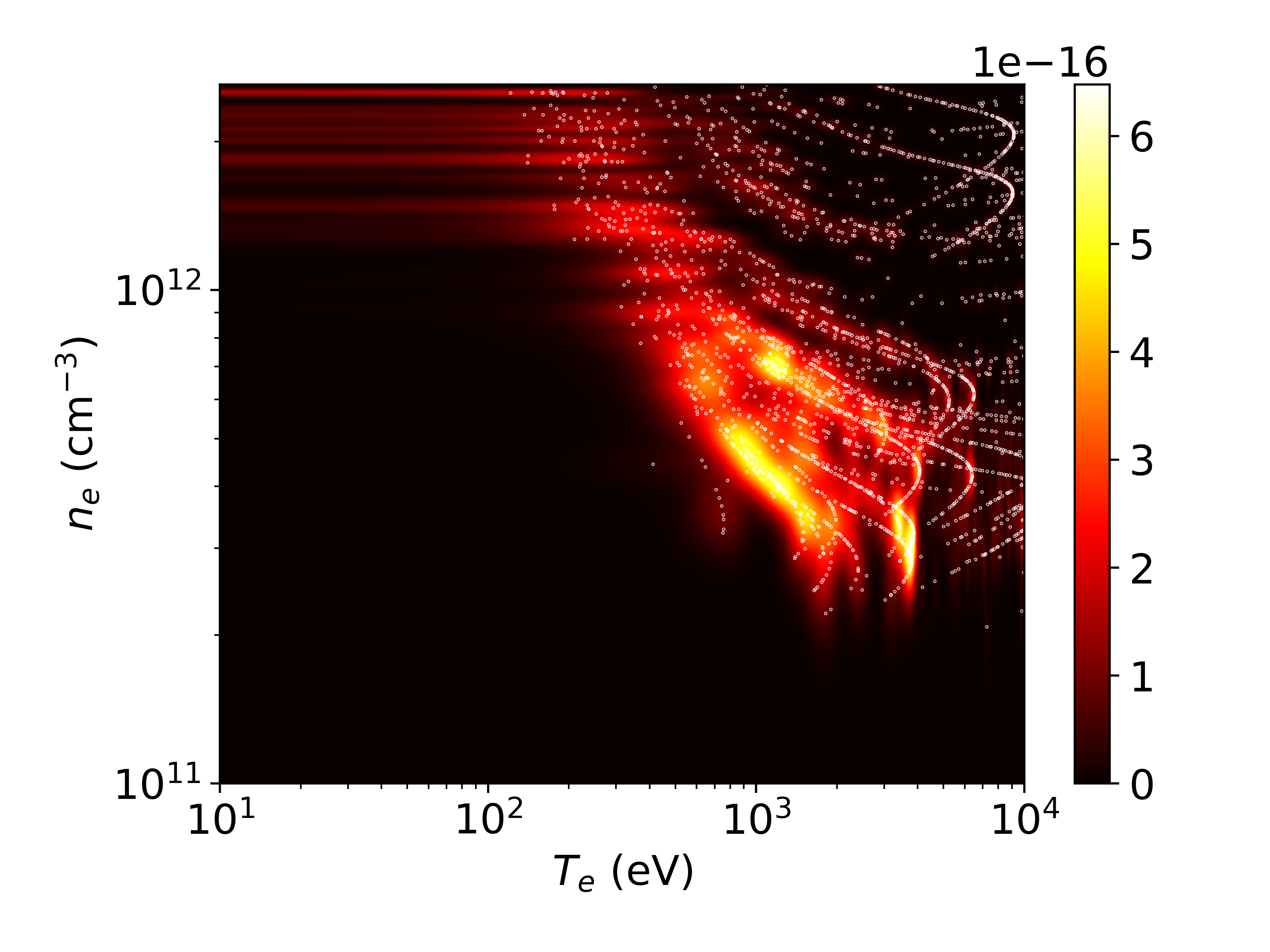}
\caption{K$^{8+}$} \label{fig:d}
\end{subfigure}
\medskip
\begin{subfigure}{0.48\textwidth}
\includegraphics[width=\linewidth]{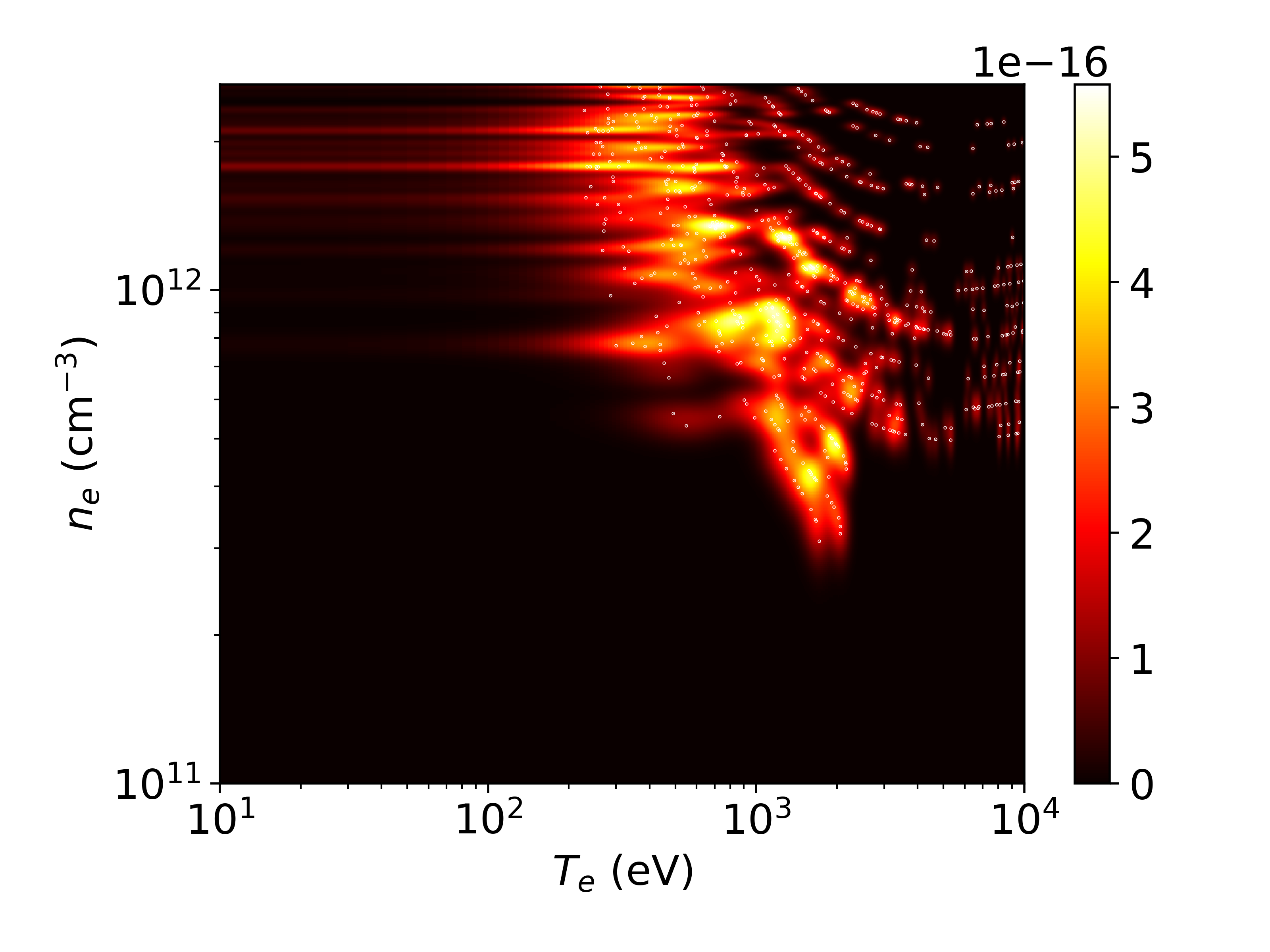}
\caption{K$^{9+}$} \label{fig:e}
\end{subfigure}\hspace*{\fill}
\begin{subfigure}{0.48\textwidth}
\includegraphics[width=\linewidth]{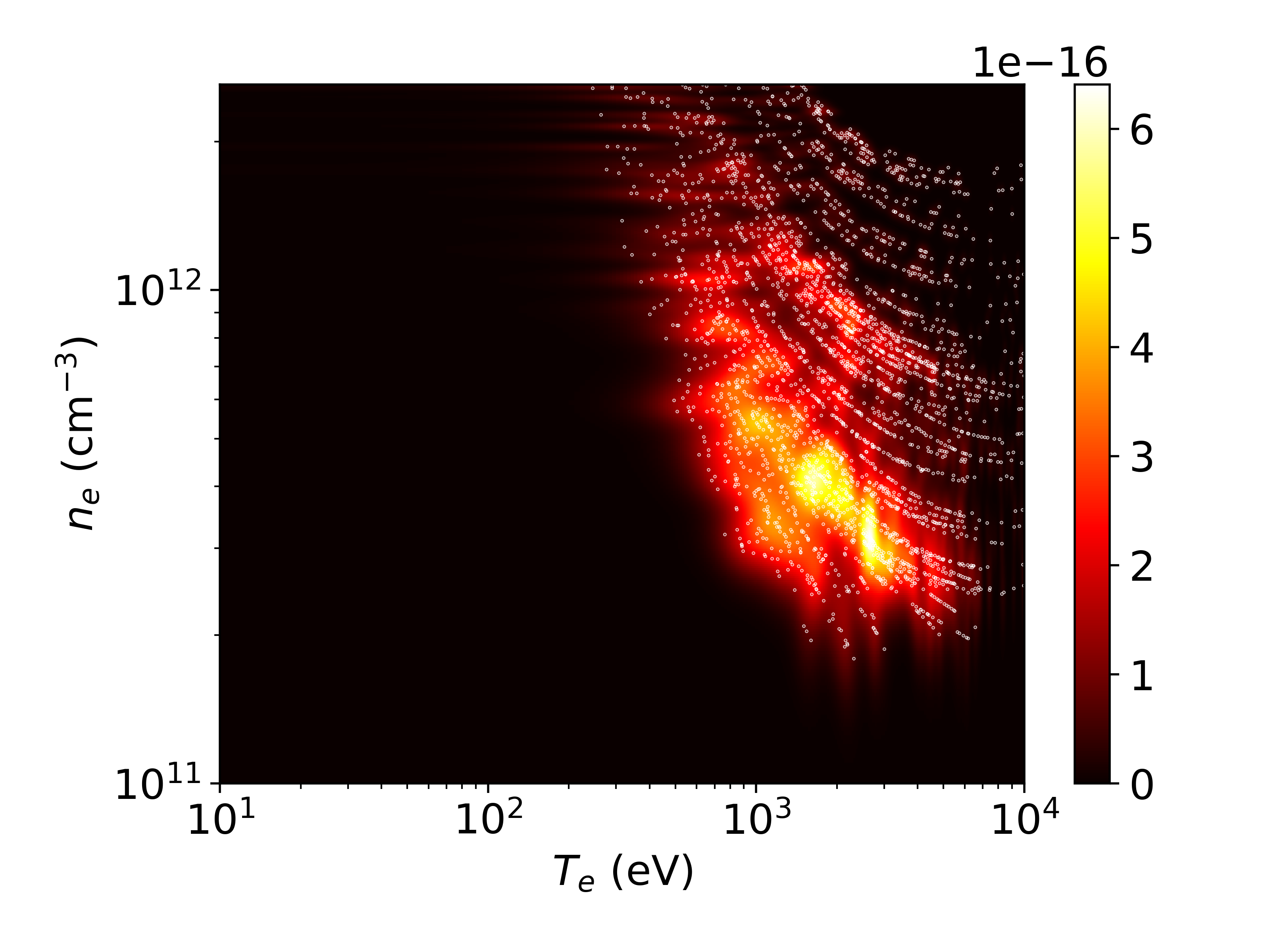}
\caption{K$^{10+}$} \label{fig:f}
\end{subfigure}

\caption{Probability density plots of the solution sets for K$^{5+}$--K$^{10+}$ (subfigures (a)--(f)) plotted as heatmaps with an applied gaussian filter. The densities are each normalized such that an integral over the figure area equals to one. The ($n_e, T_e$)-pairs of each solution set are overlaid on the heatmaps as small white dots.} \label{fig:results_solution_space}
\end{figure}

\begin{figure}[t!]
    \centering
    \begin{subfigure}[t]{0.5\textwidth}
        \centering
        \includegraphics[width=\textwidth]{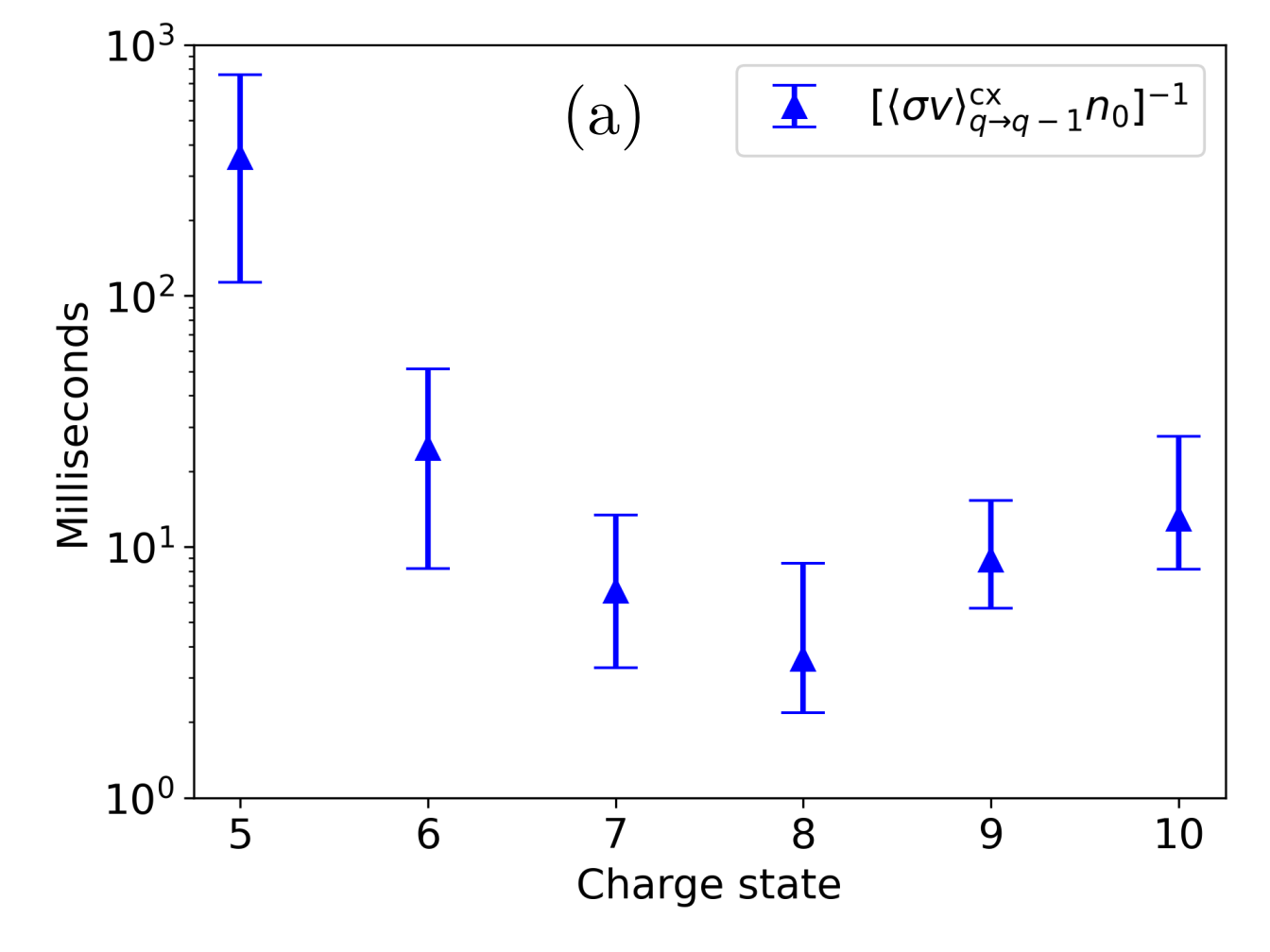}
    \end{subfigure}%
    ~ 
    \begin{subfigure}[t]{0.5\textwidth}
        \centering
        \includegraphics[width=\textwidth]{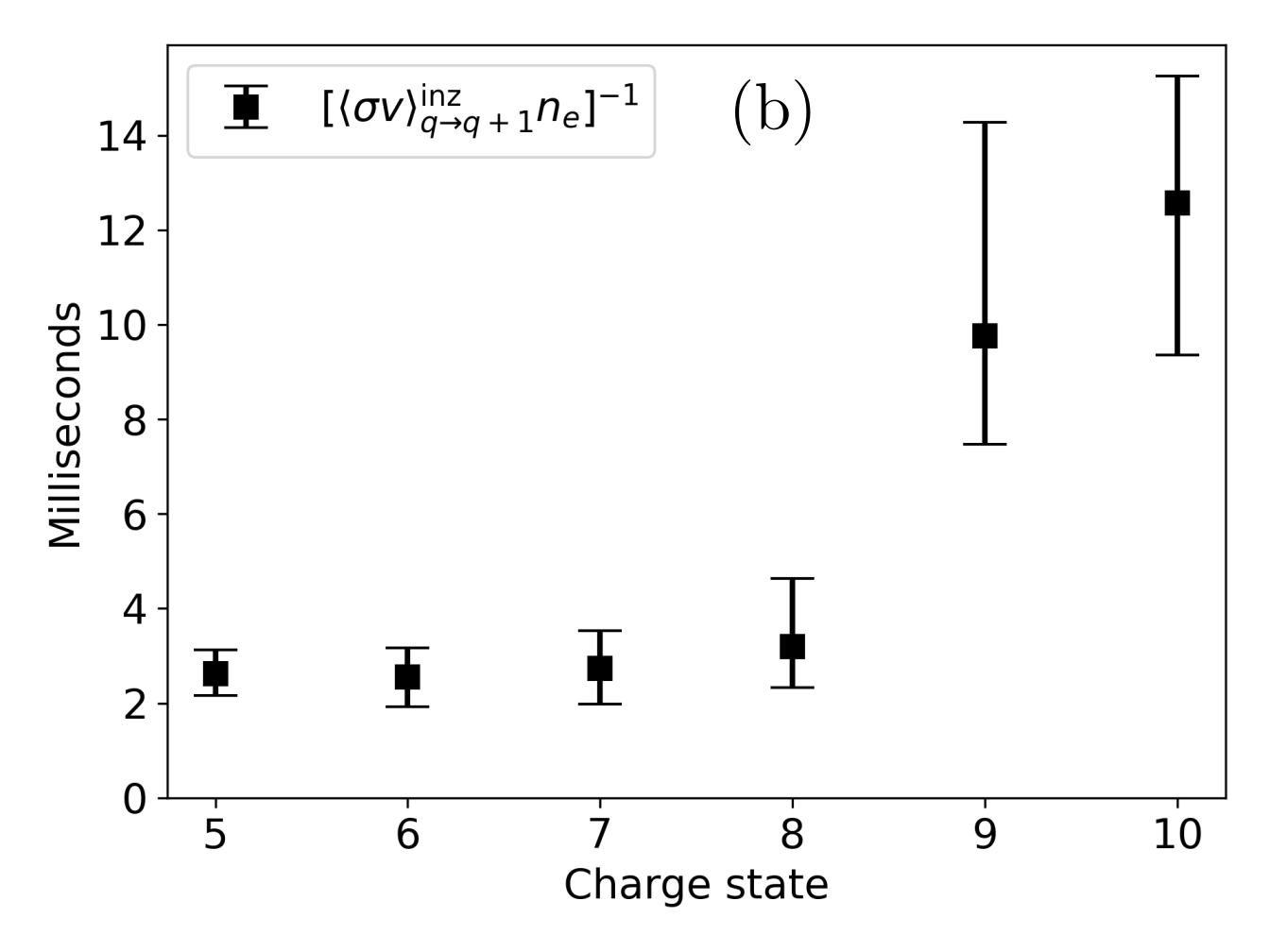}
    \end{subfigure}
    ~ 
    \begin{subfigure}[t]{0.5\textwidth}
        \centering
        \includegraphics[width=\textwidth]{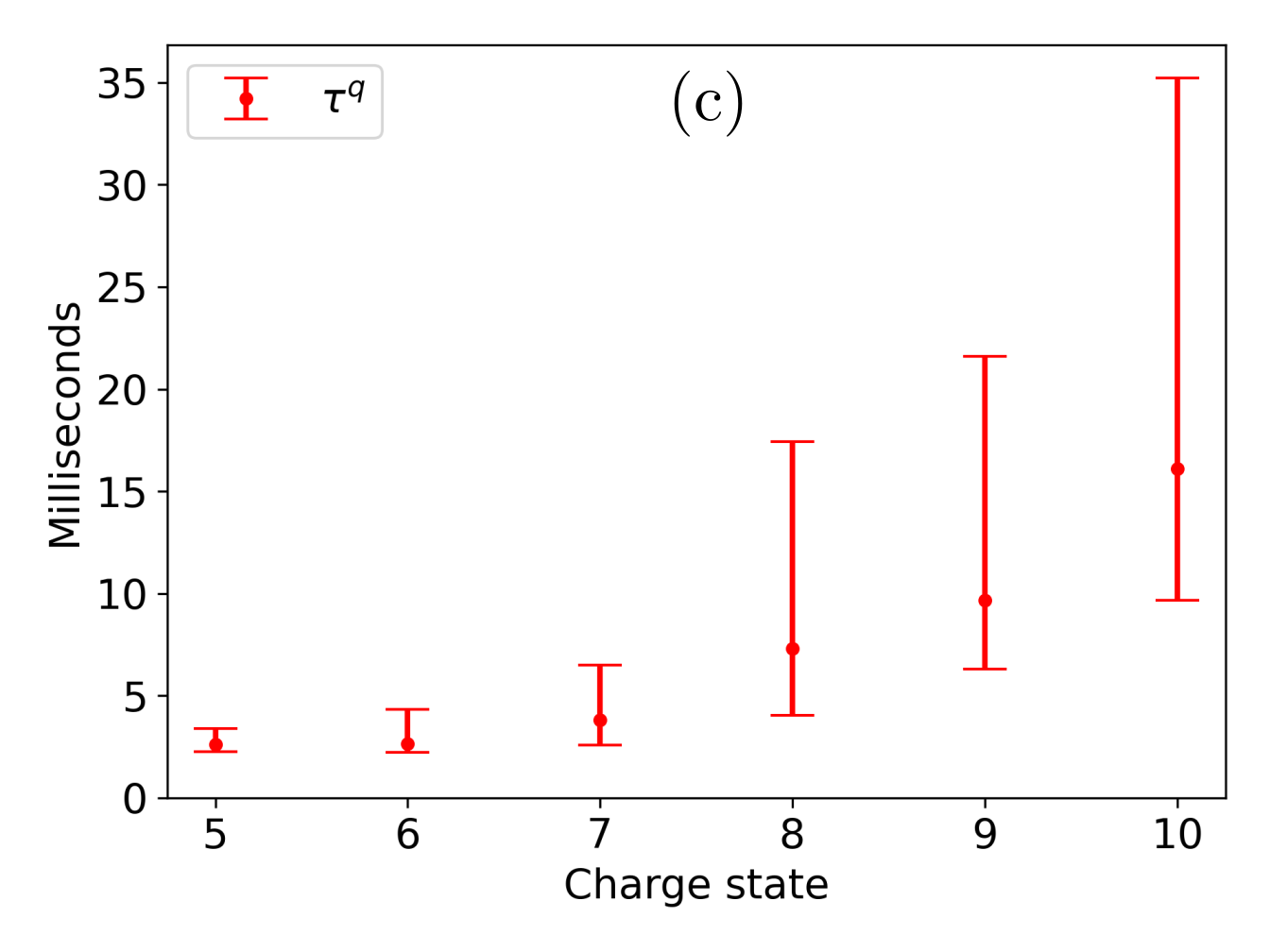}
    \end{subfigure}
    \caption{The plasma charge exchange (a), ionisation (b) and confinement times (c) as a function of charge state. Note that subfigure (a) is plotted in logscale.}\label{fig:result_characteristic_times}
\end{figure}

\begin{figure}
    \centering
    \includegraphics[width=0.5\textwidth]{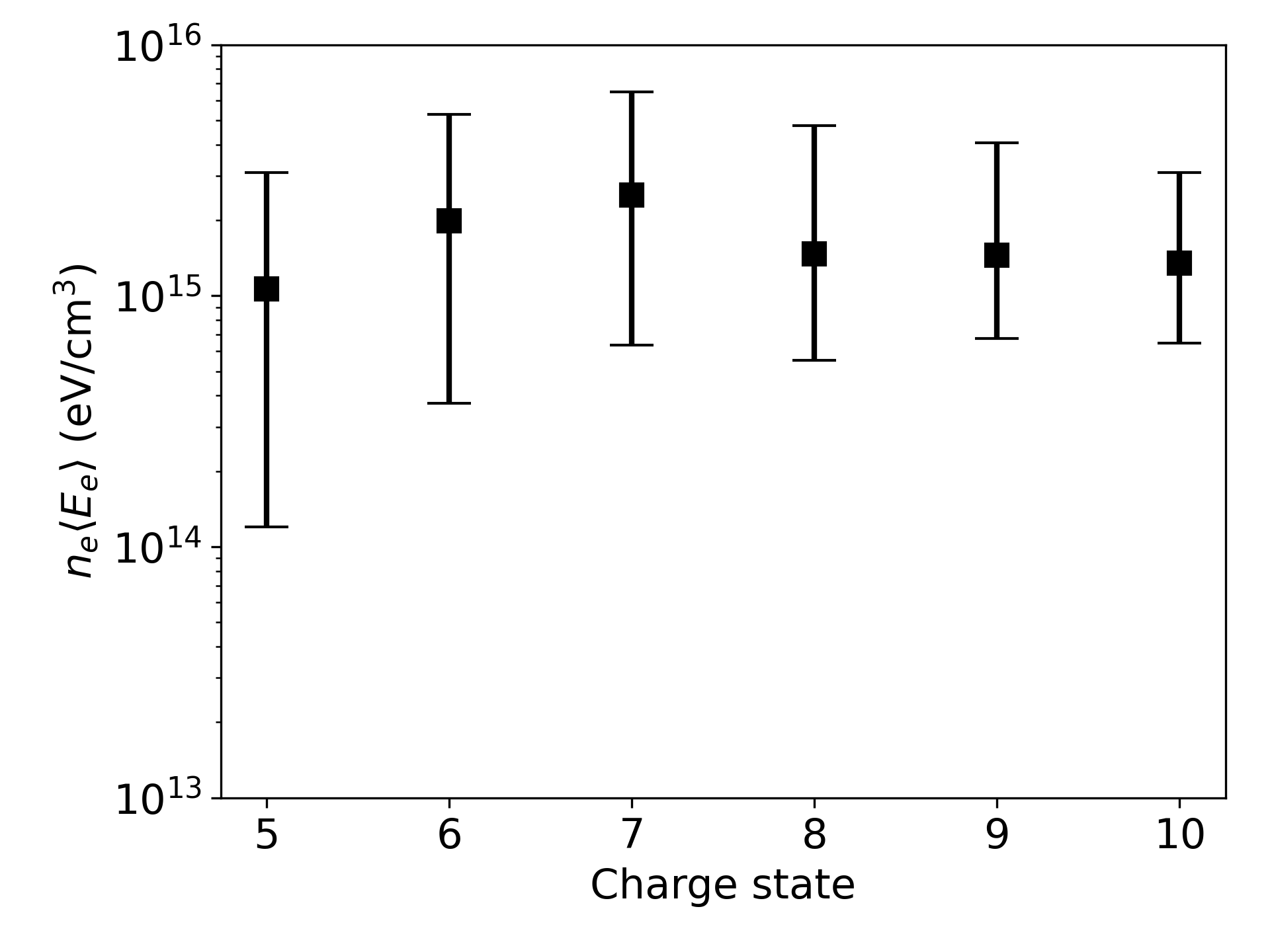}
    \caption{The local plasma energy content as a function of charge state.}
    \label{fig:result_energy_content}
\end{figure}

\begin{figure}
    \centering
    \includegraphics[width=0.5\textwidth]{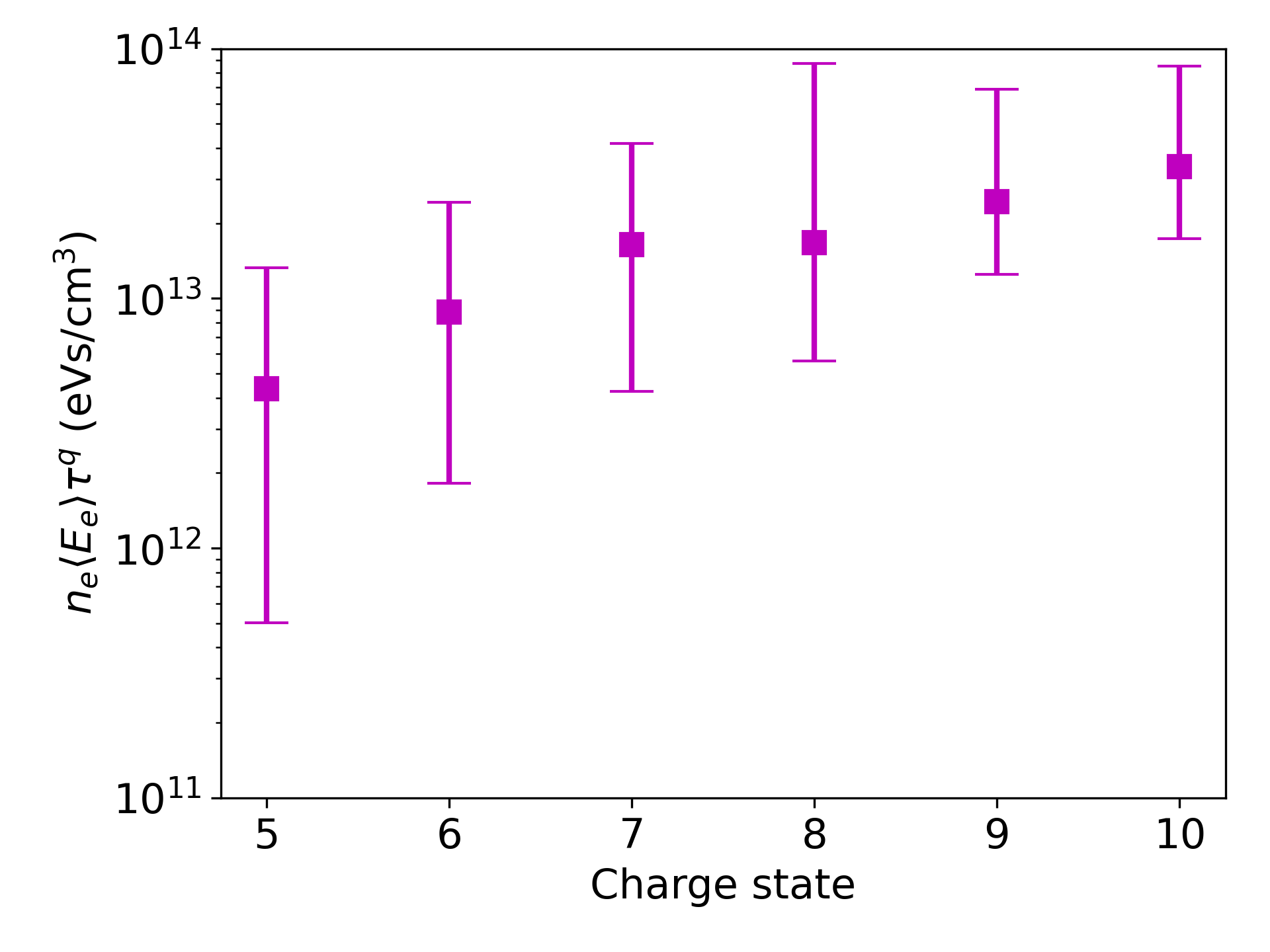}
    \caption{The local plasma triple product as a function of charge state.}
    \label{fig:result_triple_product}
\end{figure}

Figure~\ref{fig:results_solution_space} displays the sets of ($n_e, T_e$)-pairs found to satisfy Equation~\eqref{eq:eqn_of_ne_Te} for charge states K$^{5+}$--K$^{10+}$ using the method detailed in Section~\ref{sec:determining_energy_content}. Because Equation~\eqref{eq:eqn_of_ne_Te} has two unknowns, no singular solution can be found, and a set of solutions is obtained instead. The optimisation routine was run $N = 1000$ times for each charge state (in order to account for the upto 60~$\%$ uncertainty in the ionisation rate coefficients), with an optimum $T_e$ searched for 1000 $n_e$ values in the given range within each iteration. The plasma confinement, ionisation and charge exchange times ($\tau^q, \left[n_e\left\langle\sigma v\right\rangle^{\text{inz}}_{q\to q+1}\right]^{-1}$, $\left[n_0\left\langle\sigma v\right\rangle^{\text{cx}}_{q\to q-1}\right]^{-1}$), were calculated, taking into account the uncertainties issued to the rate coefficients in each iteration. The corresponding energy contents ($n_e\left\langle E_e\right\rangle$), and plasma triple products ($n_e \left\langle E_e\right\rangle \tau^q$) were similarly calculated. It should be noted that because the optimisation is performed for each charge state separately, all values obtained are local and correspond to the plasma volume relevant for the production of the charge state in question. The median values of the characteristic times, energy contents and triple products are plotted in Figures~\ref{fig:result_characteristic_times}, \ref{fig:result_energy_content} and~\ref{fig:result_triple_product} respectively. In the figures, the errorbars represent the range within which lay 34.1~\% of solutions below and above the median value; i.e. the error bars enclose a total of 68.2~\% out of all results.

The $\tau^q$ values increase as a function of charge state. This is in accordance with the trend found in Ref.~\cite{douysset00}, although the high charge state confinement times found herein are significantly longer than in their work. For example $\tau^{5+} = 2.6^{+0.8}_{-0.4}$~ms and $\tau^{7+} = 4.0^{+3.1}_{-1.4}$~ms, while $\tau^{8+} = 7.3^{+10.9}_{-3.2}$~ms, and $\tau^{10+} = 16.4^{+18.3}_{-6.8}$~ms. The high charge states are believed to be electrostatically rather than magnetically confined as their collision frequency may exceed their larmor frequency as implied by results for sodium in Ref.~\cite{tarvainen16}, and oxygen in Ref.~\cite{tarvainen15}. The ionisation time is level up to charge state 8+ (e.g. $\left[n_e\left\langle\sigma v\right\rangle^{\text{inz}}_{5+\to 6+}\right]^{-1} = 2.6^{+0.5}_{-0.5}$~ms and $\left[n_e\left\langle\sigma v\right\rangle^{\text{inz}}_{7+\to 8+}\right]^{-1} = 2.6^{+0.9}_{-0.8}$~ms), but exhibits a kink between $\left[n_e\left\langle\sigma v\right\rangle^{\text{inz}}_{8+\to 9+}\right]^{-1} = 3.1^{+1.4}_{-0.9}$~ms and $\left[n_e\left\langle\sigma v\right\rangle^{\text{inz}}_{9+\to 10+}\right]^{-1} = 9.6^{+4.0}_{-2.1}$~ms. This kink corresponds to a shell closure in the electron configuration of potassium, which is also indicated in Table~\ref{tab:voronov_coefficients}. A relatively large uncertainty bound is associated with the charge exchange times. For all except K$^{5+}$ and K$^{6+}$ the median charge exchange times were around 10~ms. 

The plasma energy content appears to have no clear charge state dependence, with all values lying around $10^{15}~$eV/cm$^3$, the average being 2.5$^{+4.3}_{-1.8}\times 10^{15}$~eV/cm$^3$. The triple product on the other hand increases with charge state until a possible saturation around K$^{10+}$. This charge dependence originates from the corresponding dependence of the confinement times.

%
%\begin{figure}
%    \centering
%    \includegraphics[width=0.5\textwidth]{fig_solution_sets.png}
%    \caption{The sets of viable solutions of Eq.~\eqref{eq:eqn_of_ne_Te} for different charge states.}
%    \label{fig:results_solution_space}
%\end{figure}
%


\section{Discussion}\label{sec:discussion}
\subsection{Method assumptions}
In this paper we have presented a new method for calculating plasma characteristic times, along with the local plasma energy contents and the triple product, from the extracted beam current transients of at least five neighboring charge state ions. The method has been applied to current transients of K$^{4+}$~--~K$^{12+}$ (see Fig.~\ref{fig:N+ responses}) obtained from short pulse injection of K$^+$ into a CB-ECRIS. 

The advantage of this method is that it relies only upon a small set of assumptions. We assume $(i)$ that the balance equation~\eqref{eq:balance_equation} adequately describes the charge state temporal evolution. Built in to the balance equation is the assumption that the ionisation process is stepwise electron-impact ionisation, and charge exchange takes place predominantly with the neutral atoms $n_0$. We also make the assumption that $(ii)$ the perturbation caused by the injected species is sufficiently small for the parameters $n_e, T_e,$ and $n_0$ to be defined by the support plasma. We further assume that $(iii)$ the extracted currents and particle densities are related by Equation~\eqref{eq:current_by_west}, where the beamline transmission coefficient $\kappa$ is assumed to be the same for any three consecutive charge states. These assumptions allow the conversion of the balance equation to the extraction current formalism, yielding equation~\eqref{eq:balance_eqn_current}. Finally, the use of Eq.~\eqref{eq:balance_eqn_current} to make fits to the extracted N+ current transients requires us to assume, that $(iv)$ $n_e, T_e, n_0$ and $\tau^q$ are constants --- or at least vary only slowly --- in time. 

Assumption $(i)$ essentially requires, that all other processes except for confinement losses, the stepwise ionisation process, and charge exchange contributing to the CSD time evolution are negligible. This means in effect that wall recycling, secondary ionisation and radiative recombination can be neglected. Using Ar here as a proxy for K --- the two elements being neighbours on the elemental table and K$^{1+}$ having the same electronic shell configuration as neutral Ar --- according to experimental data, for example the double ionisation process Ar$^{3+} + e^- \to$ Ar$^{5+}$ + 3e$^-$ is roughly an order of magnitude less likely than the single ionisation process Ar$^{3+} + e^- \to$ Ar$^{4+}$ + 2e$^-$~\cite{tawara87}, while for higher charge states the difference is even greater. Wall recycling is minimised thanks to potassium being an alkali. Charge exchange with neutrals can be seen to be much more probable than with higher charge state ions, as the cross section decreases according to the inverse square of the ionisation potential --- as evident from Eq.~\eqref{eq:cross_section_cx}, and because the low ion temperature does not allow two ions to come close enough to each other for them to exchange charge. Radiative recombination is neglected as a relatively inconsequential process as argued by Mironov \textit{et al.} in Ref.~\cite{mironov15} based on the results from Ref.~\cite{chung05}. As noted in section~\ref{sec:experimental_setup} we are also safe regarding assumption $(ii)$ since the total effect of the K$^+$ beam injected into the source in CW mode causes a minor (at most $\lesssim 5.4$~\% for the oxygen impurity) modification on the support plasma CSD. The total of $2.2\times10^{10}$ particles injected during the 5~ms pulse into the entire plasma volume constitutes a very minor perturbance compared to the buffer plasma density. As regards to assumption $(iii)$ Eq.~\eqref{eq:current_by_west} is textbook material~\cite{geller_ecris}. Corresponding equations are used for example by Douysset \textit{et al.} in Ref.~\cite{douysset00} and Melin \textit{et al.} in Ref.~\cite{melin99}. The parameters $S$, $L$ and $\kappa$ in Eq.~\eqref{eq:current_by_west} are assumed to be the same for any three consecutive charge states. The transmission coefficient $\kappa$ varies slowly with the charge state, typical transmission of the N+ beamline being around 80~\%~\cite{angot_2012}. If, however, $S$, $L$, and $\kappa$ could be determined for each charge state, they could also be included in the calculations. Defining them for all charge states for the purposes of this method paper was deemed impractical. In the final assumption $(iv)$ the constancy of $n_e, T_e$, and $n_0$ is justified as the perturbation caused by the $K^+$ pulse is indeed minimal (c.f. assumption $(ii)$).\footnote{The fitting range analysis provided in the supplementary material (see supplement~\ref{supplement:fitting_range}) also shows that the fitting parameters $a_q$, $b_q$ and $c_q$ show no time dependence for the charge states 4+ onward, which is in agreement with this assumption.}

With regards to the constancy of $\tau^q$ in time, it is known that the ion temperature affects the ion confinement time. According to Ref.~\cite{huba16} the thermal equilibration time of a test particle $\alpha$ streaming into a field of particles $\beta$ is characterised by the collision frequency
\begin{equation}\label{eq:rate_thermal_equilibration}
    \left(\tau^{\alpha/\beta}_\epsilon\right)^{-1} = \nu_{\epsilon}^{\alpha / \beta} = 1.8\cdot 10^{-19}\frac{\sqrt{m_\alpha m_\beta}q_\alpha^2 q_\beta^2 n_\beta \ln\Lambda }{(m_\alpha T_\beta + m_\beta T_\alpha)^{3/2}} \qquad (\text{s}^{-1}),
\end{equation}
where $m_{\alpha/\beta}$ is mass (g) $q_{\alpha/\beta}$ is the charge state, $n_\beta$ is particle density (1/cm$^3$), $T_{\alpha/\beta}$ is temperature (eV), and the subscripts $\alpha$ and $\beta$ refer to the test particle and field particle respectively. Table~\ref{tab:thermal_equilibration_times} tabulates the values of $\tau_{\epsilon}^{\alpha/\beta}$ for different charge states of potassium in a field of either electrons or helium ions; the experimentally obtained confinement times are also tabulated for comparison. The calculations were done for two electron populations separately with temperatures 10~eV and 1000~eV respectively, both with density $n_e = 5\times 10^{11}$~cm$^{-3}$. For the calculation in the case of helium the sum over helium charge states was taken, allowing the substitution\footnote{We note that the often used relation (c.f. Ref.~\cite{melin99}) $\sum_q n^q q^2 = n_e \left\langle q \right\rangle$ is incorrect. The derivation of Eq.~\eqref{eq:nq2} is given in supplement~\ref{supplement:derivation_of_nq2_sum}.}
\begin{equation}\label{eq:nq2}
    \sum_q n^q_\text{He} q^2 = n_e \frac{\left\langle q^2 \right\rangle}{\left\langle q \right\rangle},
\end{equation}
where the brackets denote an average value. The CSD determined from the extracted beam currents may be used for a rough estimate of the CSD in plasma~\cite{guerra13}, and we calculate from the extracted helium currents that $\left\langle q^2\right\rangle/\left\langle q\right\rangle \approx 1.2$. The $n_e$ again was set to $5\times 10^{11}$~cm$^{-3}$. A temperature of 10~eV was chosen for He, based on ion temperatures deduced in Ref.~\cite{kronholm19}. A value of $\ln\Lambda = 10$ was used for all sets, and the test particle temperature was chosen to be 1~eV.

Table~\ref{tab:thermal_equilibration_times} shows that the thermal equilibration time between potassium ions and warm electrons is greater than the confinement time, but its thermal equilibration time with the support plasma is less than a millisecond, which means that given a small enough perturbation the potassium ions can be expected to reach thermal equilibrium with the support plasma ions in a time scale much shorter than the duration of the transient. Thus the confinement time may be taken to be a constant in time. For high charge states, however, $\tau^{\alpha/\beta}_\epsilon$ between potassium and cold electrons is on the order of or less than the confinement time, which could enable the temperature of the high charge states of potassium to evolve over the duration of the transient (if the cold electron temperature is higher than the support plasma temperature). On the other hand, in Ref.~\cite{angot18} it was found that the characteristic charge breeding times were not significantly altered by longer injection pulses, which would indicate that this effect is not significant. Nevertheless, the method will be tested in future experiments using long injection pulses, allowing the plasma to reach a new equilibrium (corresponding to continuous mode 1+ injection) after the onset of the 1+ injection. In this new quiescent state the ion temperatures have certainly equilibrated. The method will then be applied to the decaying transient onset to probe the difference between the results in these two cases.

It is also implicitly assumed that the Voronov formula~\eqref{eq:voronov_rate_coeff} is accurate within the specified uncertainty range. The formula for the rate coefficient presupposes the Maxwell-Boltzmann distribution for the electron energies, and thus affects the method postdictions for $T_e$. It is known that the energy distribution of escaped electrons is non-Maxwellian~\cite{izotov18}, and a possible future upgrade to the method could be to redefine the rate coefficient formula using different EEDFs such as the Druyvesteyn, Margenau or kappa distributions~\cite{du11}.

\begin{table}
\centering
\caption{Table of characteristic thermal equilibration times for a test particle $\alpha$ = K$^{q+}$ ($T_\alpha = 1$~eV) into a field of particles $\beta$ = e$^-$ / He. The calculation was performed for two electron populations of temperature 10~eV and 1000~eV respectively. For helium $\left\langle q^2 \right\rangle/\left\langle q \right\rangle$ was 1.2 and the temperature was chosen to be 10~eV. The value $\ln\Lambda = 10$ was used, and the calculations used the same $n_e$ of $5\times10^{11}$~cm$^{-3}$. The experimentally obtained confinement times $\tau^q$ are also tabulated.}
\label{tab:thermal_equilibration_times}
\begin{tabular}{ccccc}
\hline\hline \Bstrut\Tstrut
& \multicolumn{3}{c}{$\tau_{\epsilon}^{\alpha/\beta}$} & \\
\cline{2-4} &\Bstrut\Tstrut
$\beta$ = e$^-$ & $\beta$ = e$^-$ & $\beta$ = He &\\
$\alpha$ & $T_\beta$ = 10~eV & $T_\beta$ = 1000~eV & $T_\beta$ = 10~eV & $\tau^q$\\
\midrule
K$^{+}$   & 75.39~ms & 75.39~s & 747.11~$\mu$s & \Bstrut\Tstrut\\
K$^{2+}$  & 18.85~ms & 18.85~s & 186.78~$\mu$s & \Bstrut\Tstrut\\
K$^{3+}$  & 8.38~ms  & 8.38~s  & 83.01~$\mu$s  & \Bstrut\Tstrut\\
K$^{4+}$  & 4.71~ms  & 4.71~s  & 46.69~$\mu$s  & \Bstrut\Tstrut\\
K$^{5+}$  & 3.02~ms  & 3.02~s  & 29.88~$\mu$s  & 2.6$^{+0.8}_{-0.4}$~ms \Bstrut\Tstrut\\
K$^{6+}$  & 2.09~ms  & 2.09~s  & 20.75~$\mu$s  & 2.6$^{+1.7}_{-0.5}$~ms\Bstrut\Tstrut\\
K$^{7+}$  & 1.54~ms  & 1.54~s  & 15.25~$\mu$s  & 4.0$^{+3.1}_{-1.4}$~ms\Bstrut\Tstrut\\
K$^{8+}$  & 1.18~ms  & 1.18~s  & 11.67~$\mu$s  & 7.3$^{+10.9}_{-3.2}$~ms\Bstrut\Tstrut\\
K$^{9+}$  & 0.93~ms  & 0.93~s  & 9.22~$\mu$s   & 10.0$^{+12.8}_{-3.2}$~ms\Bstrut\Tstrut\\
K$^{10+}$ & 0.75~ms  & 0.75~s  & 7.47~$\mu$s   & 16.5$^{+18.3}_{-6.8}$~ms\Bstrut\Tstrut\\
K$^{11+}$ & 0.62~ms  & 0.62~s  & 6.17~$\mu$s   & \Bstrut\Tstrut\\
K$^{12+}$ & 0.52~ms  & 0.52~s  & 5.19~$\mu$s   & \Bstrut\Tstrut\\
\hline\hline
\end{tabular}
\end{table}

\subsection{Results discussion}
The necessary conversion of the balance equation to extraction current formalism complicates the procedure of deconvolving the plasma parameters from the measurement data, and causes it to be mathematically impossible to obtain a singular $n_e$, $T_e$ pair. We have, however, been able to show that the physically allowed $n_e, T_e$ pairs result in plasma characteristic times, energy contents and triple products within a tolerably narrow distribution around a certain median value. This method is a definite improvement over the pre-existing methods thanks to the smaller number of assumptions made in the process, and also due to it relaxing the requirement of a single, global $(n_e, T_e)$ value. To the Authors' best knowledge, this is also the first time that the uncertainty of the ionisation rate coefficients is accounted for in a 0D calculation.

The results indicate that $\tau^q$ is not simply linearly dependent on $q$. There is mounting evidence, that the high charge state ions are electrostatically, rather than magnetically confined. The electrostatic confinement model presumes that a potential dip is formed in the plasma potential profile by the well confined hot electron population. This would be in accordance with the nested-layer (or onion) model for ion production in the ECRIS plasma; In the layered view the high charge states are produced in the plasma core, with lower charge states originating from larger radial distance from the chamber axis. The layer-model is supported by numerical simulations~\cite{mironov09, mironov15}, and experimental measurements of ion beam emittance~\cite{leitner01, wutte01}: Such lower beam emittances of HCIs can only be explained by considering them to be extracted from a surface smaller than the plasma electrode aperture~\cite{mandin96}. Spatially resolved beam profile measurements~\cite{panitzsch11a, panitzsch11b} have shown that ions with a higher $q/m$ ratio are indeed extracted closer to the beamline axis. The long $\tau^q$ values of the high charge states found in this work are commensurate with this view, as they are believed to be formed and trapped in the potential dip, which is formed in the plasma core. The HCIs would reside in the potential dip formed in the plasma core until they have absorbed sufficient energy to overcome the potential barrier. This is also supported by the recent optical measurements which have found ion temperatures in the range of 5~eV--28~eV with a charge state dependence~\cite{kronholm19}. 

The necessary condition for magnetic confinement is that the gyration time $\tau_\text{gyro}$ around a field line be shorter than the mean collision time. The gyration time is defined through the cyclotron frequency $\omega_c$ such that
\begin{equation}
    \frac{1}{\tau_\text{gyro}} = \frac{\omega_c}{2\pi} = \frac{q e B}{2\pi m}
\end{equation}
Here $B$ is the field intensity, and $m$ the particle mass. Table~\ref{tab:collision-and-gyration-times} shows the characteristic collision time for certain K ions in a He support plasma, tabulated alongside the gyration time around a field line with $B = 0.5$~T --- corresponding to the 14~GHz cold electron resonance field. It can be seen, that even at low temperatures $\tau^{\text{K/He}}_\epsilon$ can exceed the gyration time for low charge state potassium ions, allowing them to be magnetically confined. Meanwhile the collisionality of the HCIs can interfere with the magnetic confinement. In Refs.~\cite{tarvainen15} and~\cite{tarvainen16} similar conclusions have been drawn based on experimental data.

\begin{table}
\centering
\caption{Characteristic collision time $\tau^\text{K/He}_\epsilon$ between potassium ions (with temperature 1~eV or 10~eV) and support plasma helium ions (having temperature 1~eV or 10~eV) and the respective gyration times $\tau_\text{gyro}$ about $B = 0.5$~T field intensity. Support plasma was assumed to have $\left\langle q^2 \right\rangle / \left\langle q \right\rangle = 1.2$ and $n_e = 5\times 10^{11}$~cm$^{-3}$.}
\label{tab:collision-and-gyration-times}
\begin{tabular}{cccc|ccc|ccc}
\hline\hline\Tstrut\Bstrut
& \multicolumn{3}{c}{$T_\text{He} = 1$~eV}  & \multicolumn{3}{c}{$T_\text{He} = 10$~eV} & \multicolumn{3}{c}{$B = 0.5$~T} \\
& \multicolumn{3}{c}{$\tau^\text{K/He}_\epsilon$~($\mu$s)}  & \multicolumn{3}{c}{$\tau^\text{K/He}_\epsilon$~($\mu$s)} & \multicolumn{3}{c}{$\tau_\text{gyro}$~($\mu$s)} \\ 
\cline{2-4} \cline{5-7} \cline{8-10}\Bstrut\Tstrut 
$T_{\text{K}^{q+}}$~(eV)  & K$^{1+}$ & K$^{5+}$   & K$^{10+}$  & K$^{1+}$ & K$^{5+}$    & K$^{10+}$  & K$^{1+}$   & K$^{5+}$   & K$^{10+}$  \\ \hline\Bstrut\Tstrut
1  & 26.9     & 1.1 & 0.3 & 747.1     & 29.9 & 6.2 & 5.2 & 1.0 & 0.5 \\
10 & 67.1     & 2.7 & 0.7 & 851.8     & 34.1 & 8.5 &     &     &    \\\hline\hline
\end{tabular}
\end{table}

In Refs~\cite{neben16_experiments, neben16_analysis, marttinen20}, the confinement time has been studied by exponential fits to the decaying current transients in long pulse material injection mode. In Ref.~\cite{marttinen20} it is argued that the time constant of the decay represents the cumulative confinement time $\tau_\mathrm{c}^q$ of an individual particle, rather than the population confinement time $\tau^q$ as defined in the balance equation, which is obtained as a result of the method introduced in this work.

Since the determination of both the cumulative confinement time, the charge breeding time and the 90~\% extraction time ($\tau_\text{CB}$, $T_{90\%}$; See e.g. Ref.~\cite{angot18}) is much simpler than carrying out the complete method proposed herein, it should be worth comparing the behavior of these time scales in parameter sweeps. If a correspondence can be established between $\tau^q_{\text{c}}$, $\tau_\text{CB}$, $T_{90\%}$ and $\tau^q$, one could perform diagnostics of $\tau^q$ for a wider variety of elements. The novel method presented herein requires it to be possible to measure multiple neighboring charge states' currents, and to be able to calculate the rate coefficients for ionisation. The former condition inhibits many gas mixing experiments due to overlapping peaks in the CSD, and the latter is only possible for some of the light elements as cross section data is scarce, and even for those elements the uncertainties on the cross section data can be considerable --- the uncertainty reported in Ref.~\cite{voronov97} for example is 40~\%--60~\%. The rate coefficient uncertainty leads to the $(n_e, T_e)$ solution sets becoming ``smeared'', and consequently, if the cross sectional data were more precise the error bars on the results found using this method could also be reduced.\footnote{It should be noted that the Voronov formula reproduces the literature values for the rate coefficients to within $\sim10~$\%, but the literature values themselves have an uncertainty of $40~\%-60~$\%.}

In Ref.~\cite{imanaka05} the balance equation for ion densities (Eq.~\eqref{eq:balance_equation}) is fit to the transients of extracted beam intensities in order to determine $\tau^q$, $n_e$ and $T_e$. The confinement time of the highest charge state ion they find to be on the order of 10~ms. They have also found $n_e$ to be approximately $5\times10^{11}$~cm$^{-3}$, and $T_e$ around 2~keV--3~keV. The approach in Ref.~\cite{imanaka05} neglects the current-to-density proportionality $I^q \propto qen^q/\tau^q$, as proposed by Ref.~\cite{west82}, and further assumes $\tau^q = \tau_{\text{max}}\cdot q/q_\text{max}$. 

The assumption of the linear $q$-dependence for $\tau^q$ is based on the findings in Ref.~\cite{douysset00}, which also indicate much shorter confinement times --- less than 4~ms for Ar$^{16+}$ and below.  In Ref.~\cite{douysset00} the confinement time was determined from the saturation currents directly, by using Equation~\eqref{eq:current_by_west}, estimating the necessary parameters, $\kappa, S, L$, and by determining the population densities $n^q$ using optical spectroscopy methods. These estimations are reported in Ref.~\cite{douysset00} to be reliable within a factor on the order of 2. It should be noted, that using the Voronov formula for the ionisation rate coefficient, the electron temperature $T_c$ found in Ref.~\cite{douysset00} for the warm electron population, and the highest $n_e$ value of $5.7\times 10^{11}$~cm$^{-3}$ found therein, one finds $\left[n_e\left\langle\sigma v\right\rangle^{\text{inz}}_{15+\to 16+}\right]^{-1} \simeq (35 \pm 21)$~ms. If this were true, then for the $\tau^{15+}_\text{Ar} < 4$~ms confinement time found in Ref.~\cite{douysset00}, there should be next to no Ar$^{16+}$ produced\footnote{Ref.~\cite{douysset00} finds two approximately Maxwellian electron populations with temperatures $T_c \simeq 10$~keV and $T_h \simeq 50$~keV, with $T_h$ increasing noticably for higher $\mu$W power. In the calculation here, $T_c$ was used as it produced the shorter (more favorable) ionisation time.}, since a criterion for appreciable ionisation from $q$ to $q+1$ is that~\cite{geller_ecris}
\begin{equation}\label{eq:inz_criterion}
    \tau^q > \left[n_e\left\langle\sigma v\right\rangle_{q\to q+1}^\text{inz}\right]^{-1}.
\end{equation}
Hence, there is reason for some doubt concerning the precision of the measurements in Ref.~\cite{douysset00}.

The present method also produces estimates for the ionisation and charge exchange times in the plasma. For the purposes of HCI production these are extremely important time scales. One wants the ionisation to higher states to occur as quickly as possible, while the charge decreasing charge exchange process should ideally never occur. The lower charge state ions should be confined long enough for them to be ionized to higher states, while the desired HCIs should ideally be immediately ejected to avoid ionisation and charge exchange related losses.

The obtained ionisation times are in agreement with the behavior of the ionisation potential of potassium (tabulated in Table~\ref{tab:voronov_coefficients}), where a shell closure inhibits the ionisation from 9+ to 10+, explaining the discrete jump in ionisation times between $\left[n_e\left\langle\sigma v\right\rangle_{8+\to 9+}^\text{inz}\right]^{-1}$ and $\left[n_e\left\langle\sigma v\right\rangle_{9+\to 10+}^\text{inz}\right]^{-1}$. It can be seen that $\tau^q$ is never smaller than the ionisation time from $q$ to $q+1$ --- satisfying requirement~\eqref{eq:inz_criterion}.

The charge exchange times are systematically longer than the ionisation times, which is to be expected as the operation of an ECRIS relies on minimising the neutral density. Figure~\ref{fig:result_characteristic_times} shows that charge exchange may begin to inhibit the production of K$^{9+}$ for example. It is likely that in the case of a helium support plasma charge exchange plays a lesser role than it would e.g. if an oxygen support were used, due to the higher ionisation energy of neutral helium (13.62~eV and 24.59~eV respectively). In supplement~\ref{supplement:neutral_density} the neutral density was deconvoluted from the charge exchange times as a function of the ion temperature, and values on the order of 10$^{12}~$cm$^{-3}$ were found.\\

The plasma energy content has also been obtained as a function of charge state. Figure~\ref{fig:result_energy_content} indicates that the energy content is more or less the same throughout the plasma, which is in line with the observed increase in ionisation time, as the flat energy content cannot compensate for the increasing ionisation energy required for the production of higher charge states.\\

The plasma triple product shown in Figure~\ref{fig:result_triple_product} shows that $n_e\left\langle E_e \right\rangle \tau^q$ increases as a function of $q$. It is ``wisdom of the trade'' that a higher triple product enables production of higher charge states, as hinted by the famous Golovanivsky diagram (see e.g. Ref.~\cite{geller_ecris}). The flat energy content must be compensated for by an increase in the confinement time to allow HCI production with the increasing ionisation time. Parameter dependencies of the triple product could thus serve as a useful diagnostic when optimising the source for HCI production.\\

\printbibliography

\newpage

\appendix
\section{Supplementary material}\label{supplement}

\subsection{Fitting range analysis}\label{supplement:fitting_range}
\begin{figure}[t!]
    \centering
    \begin{subfigure}[t]{0.45\textwidth}
        \centering
        \includegraphics[width=\textwidth]{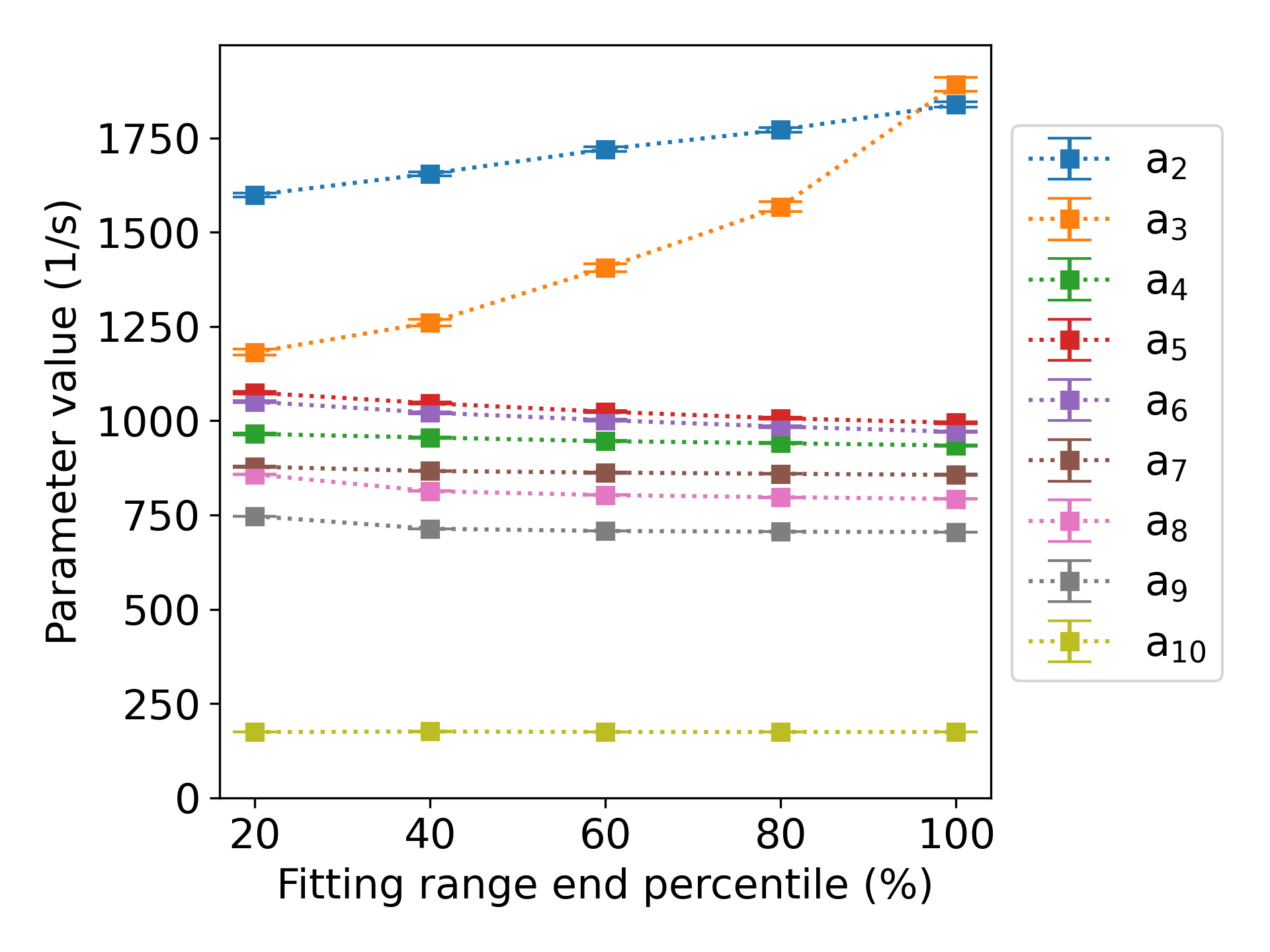}
    \end{subfigure}
    ~ 
    \begin{subfigure}[t]{0.45\textwidth}
        \centering
        \includegraphics[width=\textwidth]{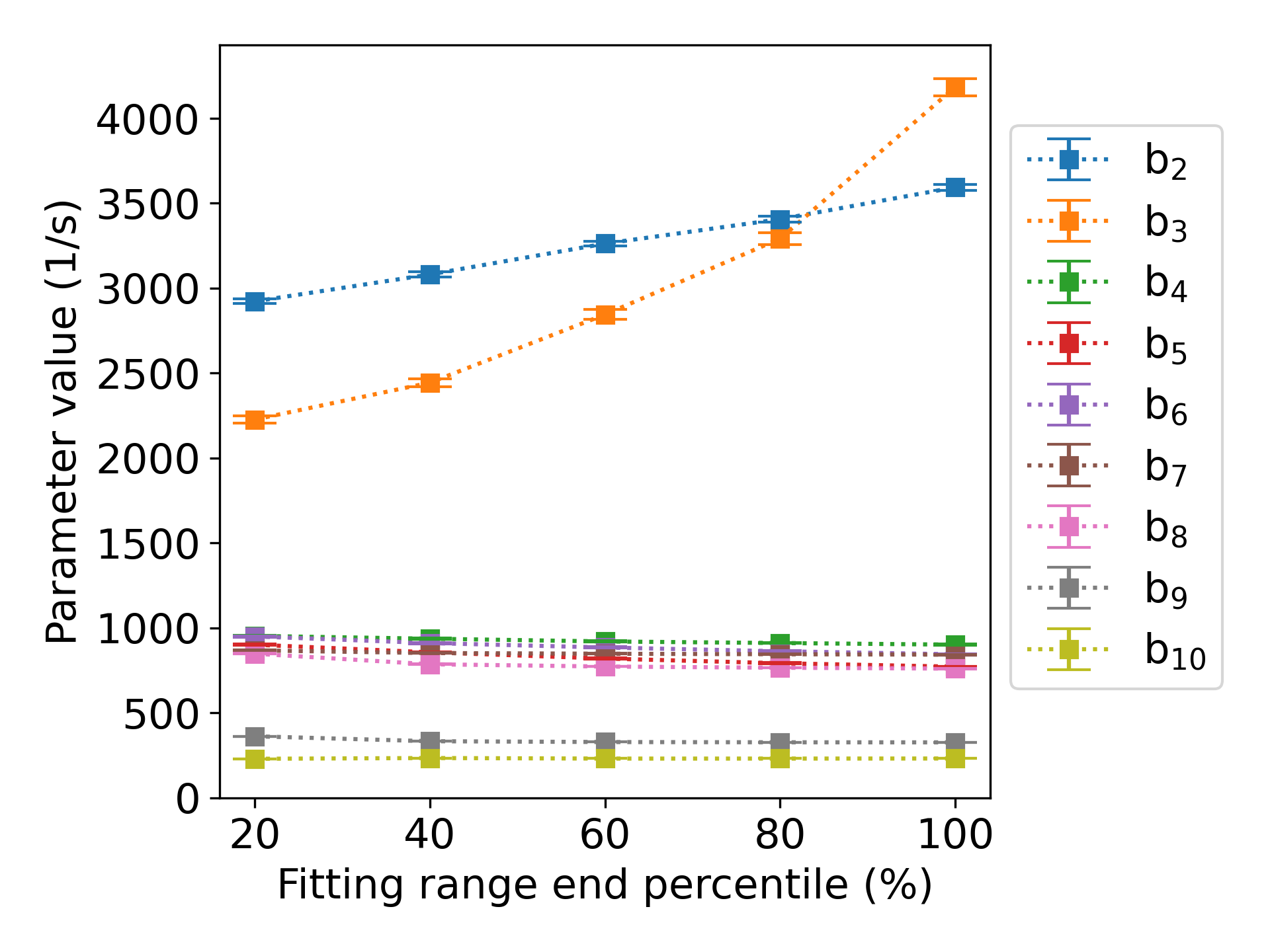}
    \end{subfigure}
    ~ 
    \begin{subfigure}[t]{0.45\textwidth}
        \centering
        \includegraphics[width=\textwidth]{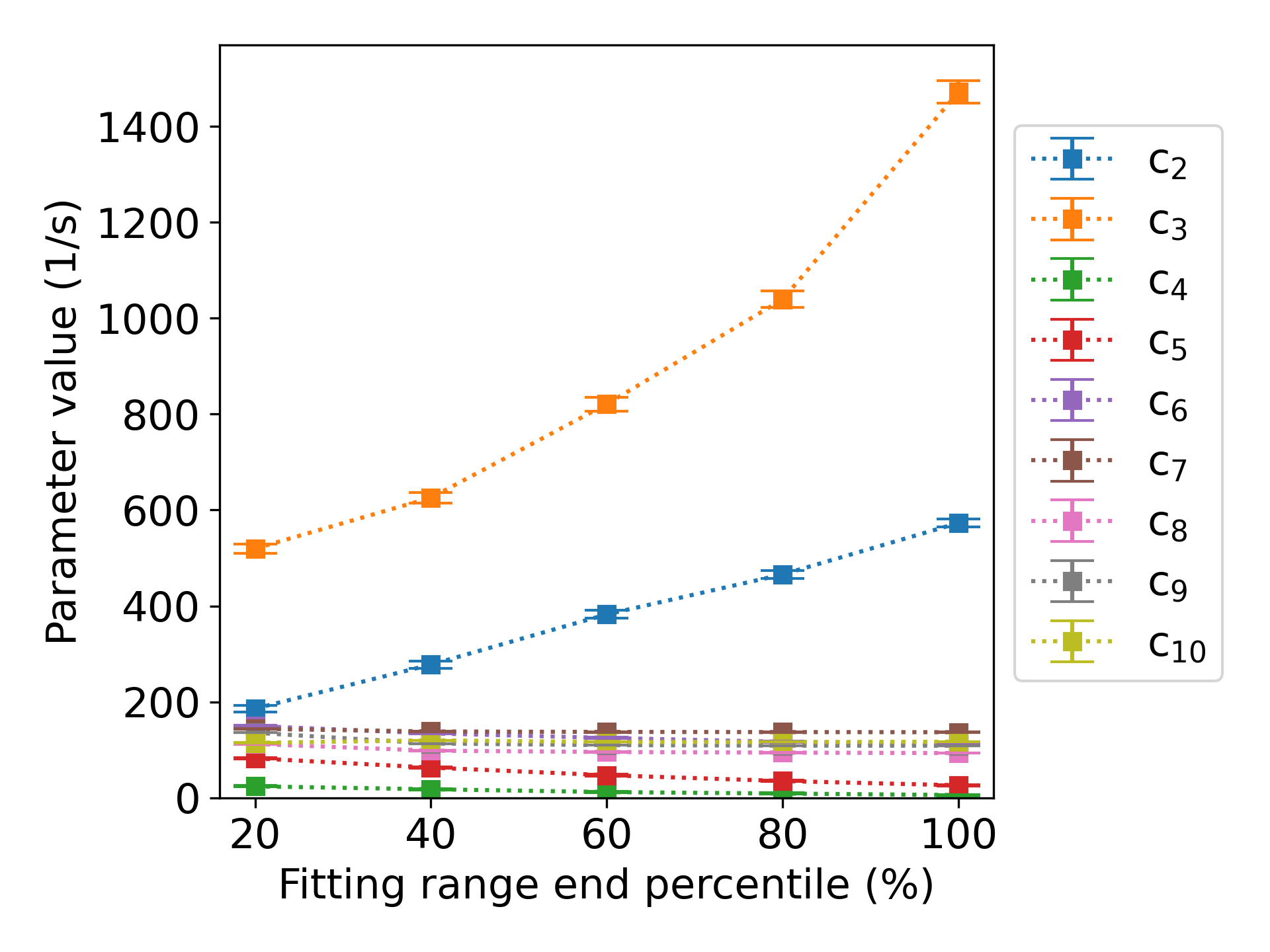}
    \end{subfigure}
    \caption{The variation of the fitting parameters $a_q$, $b_q$ and $c_q$ as a function of the fitting range.}\label{fig:params_vs_range}
\end{figure}

To study the variation in time of the fitting parameters $a_q$, $b_q$ and $c_q$, the fitting range interval was increased incrementally from $t_0 = 0$~s to $t_f$, where $t_f$ is the longest duration of an extracted transient. The fits were thus made in the intervals $\left[t_0, 0.2 t_f\right]$, $\left[t_0, 0.4 t_f\right]$, $\left[t_0, 0.6 t_f\right]$, $\left[t_0, 0.8 t_f\right]$ and $\left[t_0, t_f\right]$. Figure~\ref{fig:params_vs_range} shows the obtained fitting parameters as a function of the fitting range end percentile. It can be seen that as the fitting range is increased only the fitting parameters for charge states 2+ and 3+ change. This change is attributed to the in-flight ionised ions which are unaccounted for in the balance equation. Due to the time dependence found for the charge states 2+ and 3+, only the fitting parameters from 4+ onward were used in the analysis. 

\subsection{Numerical code}\label{supplement:numerical_code}

The python codes for obtaining the fitting parameters $a_q$, $b_q$ and $c_q$ and for determining the $n_e$, $T_e$ solution set are provided. In addition, the experimental data used in this work is provided as a sample data set. The material is available on Github, and instructions for its use will be available there in the near future.

\subsection{Penalty function limit analysis}\label{supplement:penalty_function}

\begin{figure}
    \centering
    \includegraphics[width=0.65\textwidth]{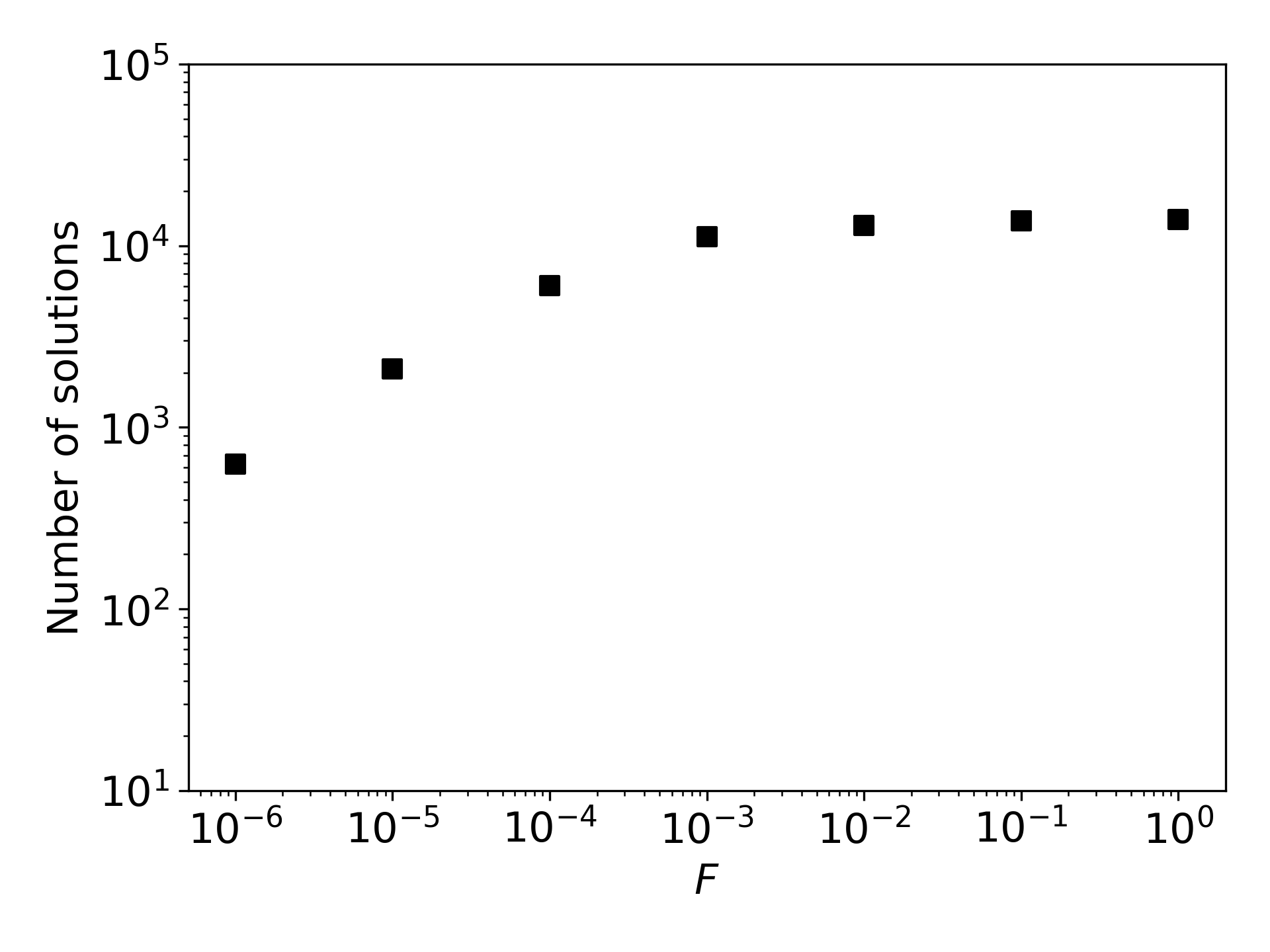}
    \caption{The number of valid solutions with a given upper limit of the penalty function $F$ for K$^{7+}$.}
    \label{fig:number_of_solutions_7+}
\end{figure}

\begin{figure}[t!]
    \centering
    \begin{subfigure}[t]{0.44\textwidth}
        \centering
        \includegraphics[width=\textwidth]{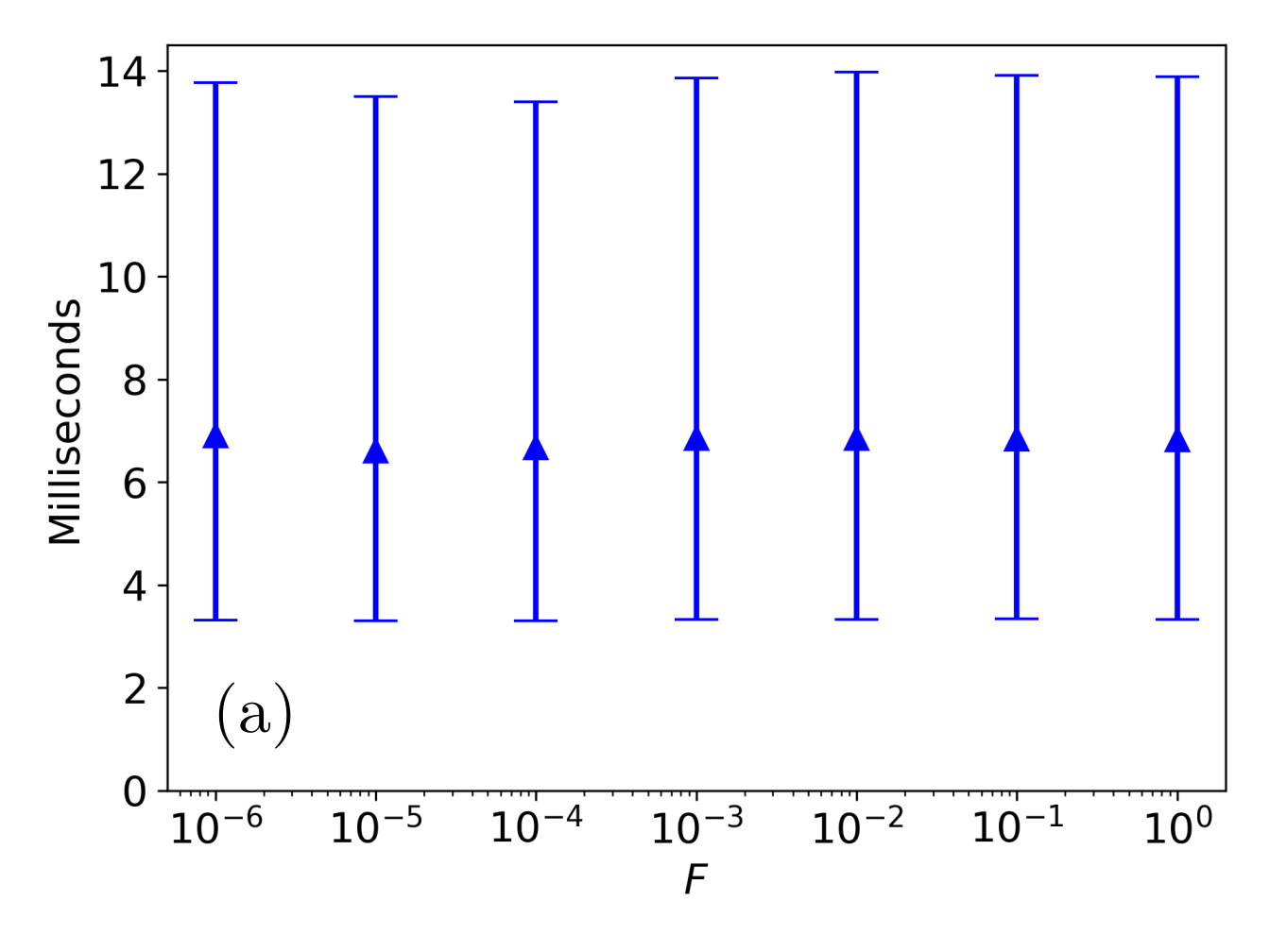}
    \end{subfigure}
    ~ 
    \begin{subfigure}[t]{0.44\textwidth}
        \centering
        \includegraphics[width=\textwidth]{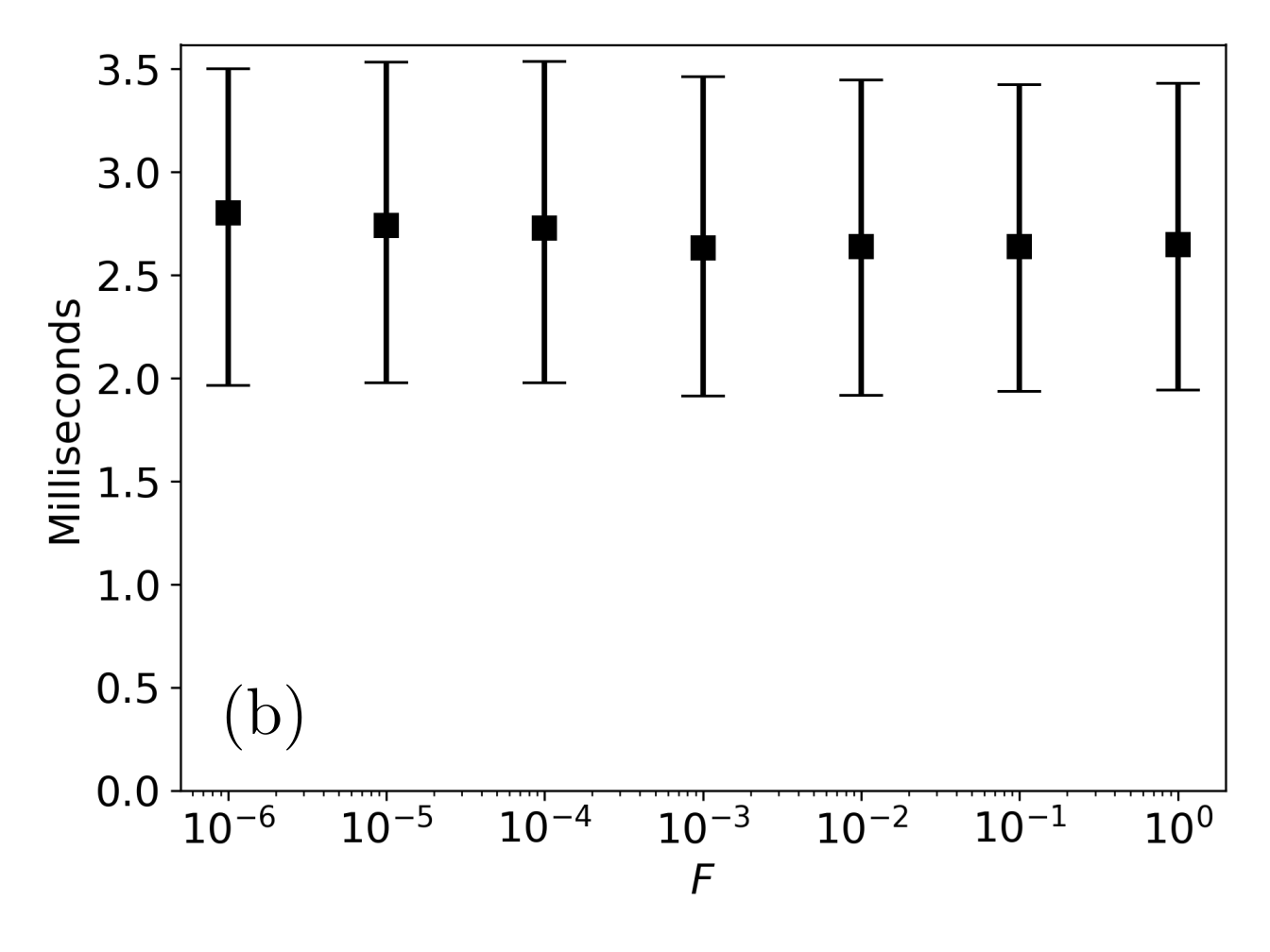}
    \end{subfigure}
    ~ 
    \begin{subfigure}[t]{0.44\textwidth}
        \centering
        \includegraphics[width=\textwidth]{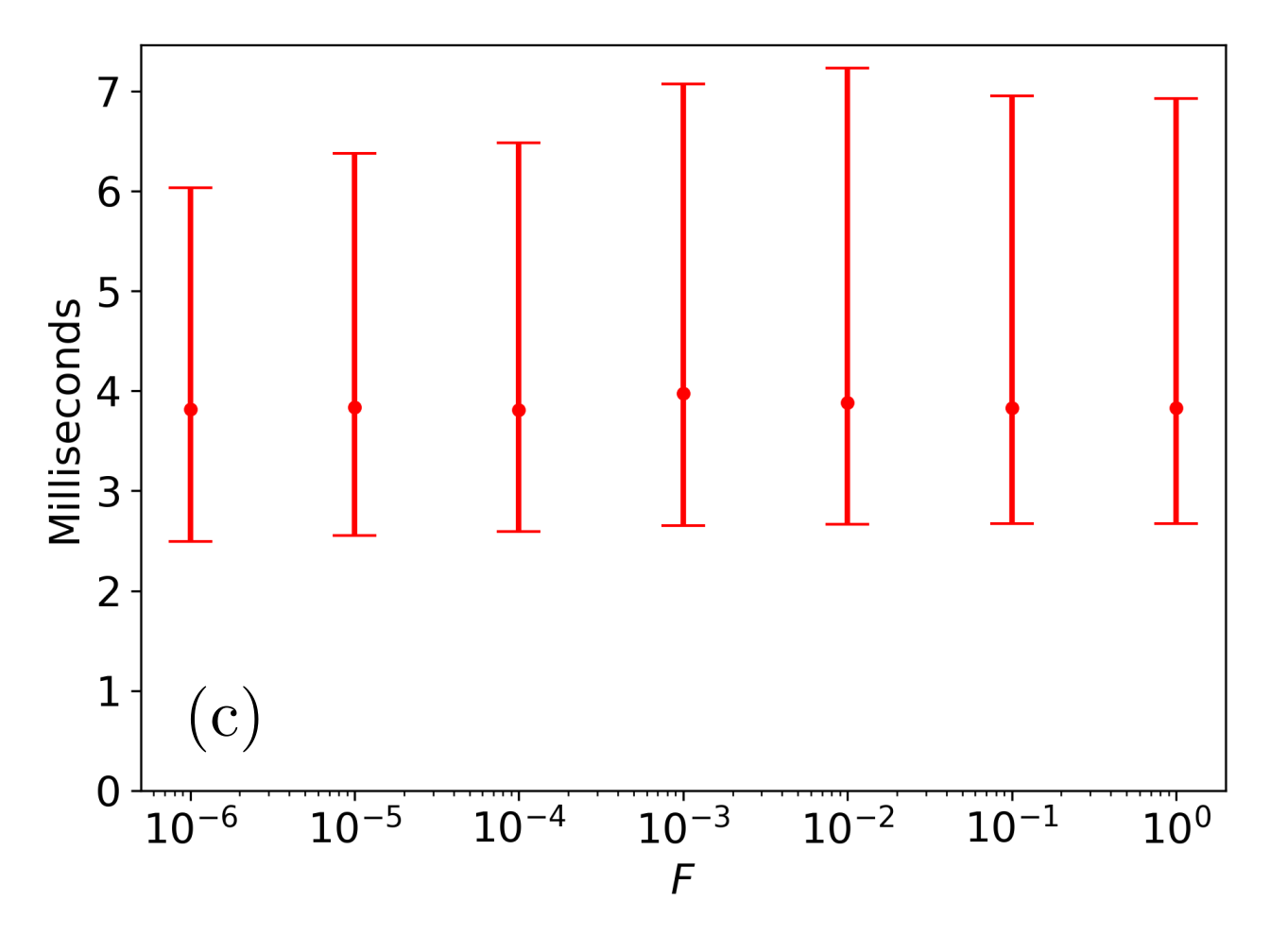}
    \end{subfigure}
    \caption{The postdictions of the plasma characteristic times for charge exchange (a), ionisation (b) and ion confinement (c) as a function of the upper limit of the penalty function $F$ for K$^{7+}$.}
    \label{fig:characteristic_times_vs_F}
\end{figure}

In order to determine the precision required by the computation presented in section~\ref{sec:numerical_method}, the effect of constraining the maximum value of the penalty function $F$ on the results obtained from the computation performed using 1000 Monte Carlo biases for the Voronov formula was studied. The smaller the maximum value of $F$, the more precisely the left-hand-side and right-hand-side of Eq.~\eqref{eq:eqn_of_ne_Te} match one another, and hence the more precise the solution. Figures~\ref{fig:number_of_solutions_7+}, and~\ref{fig:characteristic_times_vs_F} show the effect of constraining the maximum value of $F$ on the number of valid solutions and the resultant values for the plasma parameters, respectively. The results of the analysis are plotted for K$^{7+}$ as representative of all other charge states producing similar results. Based on the analysis, we choose $F < 10^{-4}$ as it is the highest precision for the solution of Eq.~\eqref{eq:eqn_of_ne_Te}, while the number of valid solutions is still some thousands for all charge states, ensuring a thorough sampling of the solution set.
\subsection{Neutral density deconvolution}\label{supplement:neutral_density}

\begin{figure}
    \centering
    \includegraphics[width=0.5\textwidth]{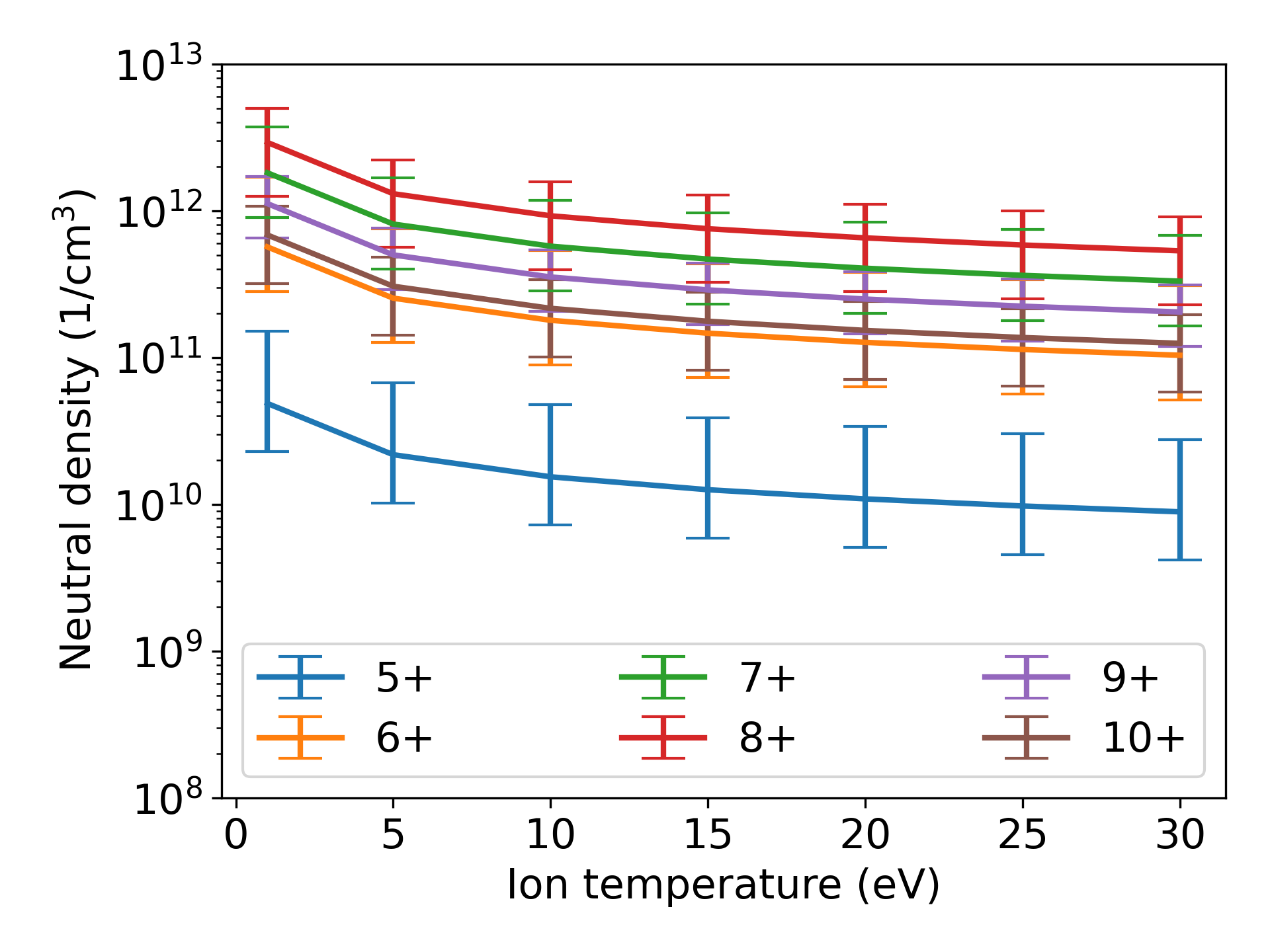}
    \caption{The postdicted local neutral density as a function of ion temperature.}
    \label{fig:n0}
\end{figure}

The charge exchange times $\left[n_0\left\langle \sigma v\right\rangle_{q\to q-1}^\text{cx}\right]^{-1}$ were obtained via the method presented in this paper. Using the equations~\eqref{eq:cross_section_cx} and~\eqref{eq:rate_coeff_cx} presented in section~\ref{sec:theory}, one can calculate the rate coefficient $\left\langle \sigma v\right\rangle_{q\to q-1}^\text{cx}$ as a function of the unknown ion temperature $T_i^q$ according to
\begin{equation}\label{eq:supp:rate_coefficient_cx}
    \left\langle \sigma v\right\rangle_{q\to q-1}^\text{cx} = \pi r_0^2 q \left(\frac{I_0}{I}\right)^2 Z_\text{eff} \sqrt{\frac{8 T_i^q}{\pi m_i}}.
\end{equation}
It should be noted that Eq.~\eqref{eq:supp:rate_coefficient_cx} assumes the neutral temperature to be low compared to the ion temperature. The effective charge state of neutral helium can be calculated by using the formulae provided in Ref.~\cite{clementi63}
\begin{equation}
    Z_\text{eff} = Z - \Sigma,
\end{equation}
where $Z$ is the proton number, and the screening coefficient $\Sigma$ for the $1s$ electronic state is given by
\begin{equation}
    \begin{split}
        \Sigma\left(1s\right) &= 0.3\times(1s - 1) + 0.0072\times(2s + 2p) \\
        &+ 0.0158\times(3s + 3p + 4s + 3d + 4p),
    \end{split}
\end{equation}
where the number of electrons in a given quantum state characterized by quantum numbers $n$ and $l$ is denoted by the terms $nl = (1s, 2s, 2p, \ldots)$. For ground state helium only the state $1s$ is occupied, and thus its effective charge is
\begin{equation}
    Z_\text{eff} = 2 - 0.3\times(2-1) = 1.7.
\end{equation}
Reference~\cite{kronholm19} has found the ion temperatures to lie in the range $T_i^q\in \left[5,28\right]~$eV. We calculate accordingly the charge exchange rate coefficient in the range $\left[1,30\right]~$eV. The neutral density can then be obtained from the characteristic frequency for charge exchange by dividing by the corresponding rate coefficient. The results as a function of ion temperature are shown in Fig.~\ref{fig:n0}.

\subsection{Collision time calculations}\label{supplement:derivation_of_nq2_sum}

The total collision frequency between the test particle $\alpha$ and particle species $\beta$ present in the plasma can be obtained by taking a sum of Eq.~\eqref{eq:rate_thermal_equilibration}. The ion-ion collision frequency between a K$^{q+}$ ion and the helium buffer is thus obtained from
\begin{equation}
    \sum_{\beta = \text{He}^+, \text{He}^{2+} } \nu^{\alpha/\beta}_\epsilon = 1.8\cdot 10^{-19}\frac{\sqrt{m_\alpha m_\beta}q_\alpha^2 \ln\Lambda }{(m_\alpha T_\beta + m_\beta T_\alpha)^{3/2}} \sum_\beta n_\beta q_\beta^2 \qquad (s)
\end{equation}
where $T_\beta$ is assumed the same for both He charge states to allow moving it out of the summation. Because the helium CSD in plasma is not precisely known an alternative form is derived as follows:

The plasma effective (average) charge state is defined according to
\begin{equation}\label{eq:definition_qeff}
    \left\langle q\right\rangle \equiv \frac{\sum_q n^q q}{\sum_q n^q}.
\end{equation}
where $n^q$ is charge state $q$ ion number density, and $\left\langle q\right\rangle$ is the effective charge state. In a multispecies plasma the summation is also carried over all ion species $i$. Due to quasi-neutrality of the plasma
\begin{equation}\label{eq:ne}
    n_e = \sum_q n^q q.
\end{equation}
The mean square charge state is defined as, 
\begin{equation}\label{eq:average_q_square}
    \left\langle q^2\right\rangle \equiv \frac{\sum_q n^q q^2}{\sum_q n^q}.
\end{equation}
From Eq.~\eqref{eq:definition_qeff} we solve
\begin{equation}\label{eq:sum_nq}
    \sum_q n^q = \frac{n_e}{\left\langle q \right\rangle},
\end{equation}
where the substitution according to Eq.~\eqref{eq:ne} was made. Substituting Eq.~\eqref{eq:sum_nq} to Eq.~\eqref{eq:average_q_square} one obtains
\begin{equation}
    \left\langle q^2 \right\rangle = \frac{\sum_q n^q q^2}{n_e}\left\langle q \right\rangle,
\end{equation}
which after rearranging gives
\begin{equation}\label{supplement:eq:correct_nq2}
    \sum_q n^q q^2 = n_e\frac{\left\langle q^2 \right\rangle}{\left\langle q \right\rangle}.
\end{equation}
The averages in Eq.~\eqref{supplement:eq:correct_nq2} can be approximated from the extracted beam currents as per
\begin{equation}
    \left\langle q^2 \right\rangle \simeq \frac{\sum_q (I^q/q) q^2}{\sum_q (I^q/q)}
\end{equation}
and
\begin{equation}
    \left\langle q \right\rangle \simeq \frac{\sum_q (I^q/q) q}{\sum_q (I^q/q)}
\end{equation}
since the extracted beam CSD provides a rough image of the CSD in plasma~\cite{guerra13}.

\end{document}